\RequirePackage{arydshln}
\documentclass[floatfix,aps,twocolumn,nofootinbib,superscriptaddress,preprintnumbers,pra,10pt]{revtex4-1}

\usepackage[utf8]{inputenc}
\usepackage{float}
\usepackage{colortbl}
\usepackage{amsmath,amssymb,amsfonts, bm,bbm,slashed}
\usepackage{dsfont} 
\usepackage{hyperref}
\usepackage{graphicx}
\usepackage{enumitem}
\usepackage{arydshln}
\usepackage{mathtools}
\usepackage{bbold}
\usepackage{braket}
\usepackage{multirow}
\usepackage{soul}

\usepackage{feynmp-auto}
\makeatletter
\g@addto@macro\bfseries{\boldmath}
\makeatother

\newcommand{\be} {\begin{equation}}
\newcommand{\ee} {\end{equation}}
\newcommand{\bea} {\begin{eqnarray}}
\newcommand{\eea} {\end{eqnarray}}
\newcommand{\no} {\nonumber}
\newcommand{\ba} {\begin{array}}
\newcommand{\ea} {\end{array}}
\newcommand{\gsim}{\lower.7ex\hbox{$\;\stackrel{\textstyle>}{\sim}\;$}}
\newcommand{\lsim}{\lower.7ex\hbox{$\;\stackrel{\textstyle<}{\sim}\;$}}
\renewcommand{\Re}{{\rm Re}}

\newcommand{\cL}{\mathcal{L}}

\newcommand{\cO}{\mathcal{O}}

\newcommand{\cB}{\mathcal{B}}
\newcommand{\cM}{\mathcal{M}}
\newcommand{\cMn}{\mathcal{M}_\chi}
\newcommand{\xZ}{x_{\Zp}}

\newcommand{\mZ}{m_{Z^\prime}}

\newcommand{\Gp}{G^\prime}
\newcommand{\Zp}{Z^\prime}

\allowdisplaybreaks

\begin{document}

\preprint{ZU-TH-32/20}

\title{Vector Leptoquarks Beyond Tree Level III:\texorpdfstring{\\}{} Vector-like Fermions and Flavor-Changing Transitions}
 
\author{Javier Fuentes-Mart\'{\i}n}
\email{fuentes@physik.uzh.ch}
\affiliation{Physik-Institut, Universit\"at Z\"urich, CH-8057 Z\"urich, Switzerland}
\author{Gino Isidori}
\email{isidori@physik.uzh.ch}
\affiliation{Physik-Institut, Universit\"at Z\"urich, CH-8057 Z\"urich, Switzerland}
\author{Matthias K\"{o}nig}
\email{matthias.koenig@uzh.ch}
\affiliation{Physik-Institut, Universit\"at Z\"urich, CH-8057 Z\"urich, Switzerland}
\author{Nud{\v z}eim Selimovi{\'c}}
\email{nudzeim@physik.uzh.ch}
\affiliation{Physik-Institut, Universit\"at Z\"urich, CH-8057 Z\"urich, Switzerland}

\begin{abstract}
\vspace{5mm}
Extending previous work on this subject, we evaluate the impact of vector-like fermions at next-to-leading order accuracy in models with a massive vector leptoquark embedded in the $SU(4)\times SU(3)^\prime\times SU(2)_L\times U(1)_X$ gauge group. Vector-like fermions induce new sources of flavor symmetry breaking, resulting in tree-level flavor-changing couplings for the leptoquark not present in the minimal version of the model. These, in turn, lead to a series of non-vanishing flavor-changing neutral-current amplitudes at the loop level.  We systematically analyze these effects in semileptonic, dipole and $\Delta F=2$ operators. The impact of these corrections in $b\to s\nu\nu$ and $b\to c\tau\nu$ observables are discussed in detail. In particular, we show that, in the parameter region providing a good fit to the $B$-physics anomalies, the model predicts a $10\%$ to $50\%$ enhancement of $\cB(B\to K^{(*)}\nu\nu)$.
\vspace{3mm}
\end{abstract}

\maketitle

\section{Introduction  \label{sec:intro}}

The $B$-physics anomalies have triggered a renewed interest in theory and phenomenology of models containing leptoquark fields. In particular, the $U_1$ massive vector leptoquark (LQ), originally proposed by Pati and Salam (PS) in the context of a unified description of quarks and leptons~\cite{Pati:1974yy}, has the correct quantum numbers to provide a successful phenomenological description~\cite{Alonso:2015sja,Calibbi:2015kma,Barbieri:2015yvd,Buttazzo:2017ixm,Bhattacharya:2016mcc,Kumar:2018kmr,Crivellin:2018yvo} of the recent anomalies (see e.g.~\cite{deSimone:2020kwi} for a recent review). A key ingredient to achieve this goal is a LQ mass around a few TeV and $\cO(1)$ couplings to third generation fermions. These requirements rule out the original PS model and have motivated the study of a series of alternative models able to host the $U_1$ field~\cite{Barbieri:2016las,Assad:2017iib,Calibbi:2017qbu,Barbieri:2017tuq,Blanke:2018sro,DiLuzio:2017vat,DiLuzio:2018zxy,Bordone:2017bld,Greljo:2018tuh,Bordone:2018nbg,Cornella:2019hct,Fuentes-Martin:2020bnh, Guadagnoli:2020tlx,Fornal:2018dqn}. Among them, those based on the gauge group $SU(4)\times SU(3)^\prime\times SU(2)_L \times U(1)_X$~\cite{DiLuzio:2017vat,DiLuzio:2018zxy,Bordone:2017bld,Greljo:2018tuh,Bordone:2018nbg,Cornella:2019hct,Fuentes-Martin:2020bnh,Guadagnoli:2020tlx} (originally proposed in~\cite{Georgi:2016xhm,Diaz:2017lit}, and denoted as ``4321'' in the following) are particularly interesting and well motivated. This is the case especially for those implementations where the SM-like fermions are charged non-universally~\cite{Bordone:2017bld,Greljo:2018tuh,Bordone:2018nbg,Cornella:2019hct,Fuentes-Martin:2020bnh,Guadagnoli:2020tlx}. The interest in such class of models goes beyond their phenomenological impact in $B$-physics: they hint at a possible solution of the Standard Model (SM) flavor puzzle~\cite{Bordone:2017bld}, and might also be able to address the electroweak hierarchy problem~\cite{Fuentes-Martin:2020bnh}.

As pointed out in~\cite{Fuentes-Martin:2019ign,Fuentes-Martin:2020luw}, in order to investigate the interplay between precision measurements and collider searches in this class of models, it is important to explore the relation between low- and high-energy observables beyond the tree level. In~\cite{Fuentes-Martin:2019ign,Fuentes-Martin:2020luw} we have presented a systematic analysis of the next-to-leading-order (NLO) corrections induced by the two largest gauge couplings, namely $\alpha_4$ and $\alpha_s$. Such NLO effects lead to a sizable enhancement of the LQ contribution in low-energy semileptonic observables, at fixed on-shell coupling, that could reach up to $40\%$ in specific amplitudes~\cite{Fuentes-Martin:2019ign}. 

The analysis of~\cite{Fuentes-Martin:2019ign,Fuentes-Martin:2020luw}, being focused on NLO effects related to the gauge sector, has been performed in a simplified version of 4321 models characterized by the minimal fermion and scalar field content. The purpose of this paper
is to go beyond this limitation by analyzing the impact of one-loop corrections due to the exchange of massive vector-like fermions. The latter are a key ingredient for a successful description of the anomalies, and also a necessary ingredient to describe the subleading entries in the effective Yukawa couplings of the SM-like chiral fermions~\cite{Bordone:2017bld,Greljo:2018tuh,Bordone:2018nbg,DiLuzio:2018zxy,Cornella:2019hct}.

More precisely, the purpose of the paper is twofold. On the one hand, extending the model with the inclusion of vector-like fermions, we evaluate the modifications of the leading  $\cO(\alpha_4)$ corrections to the matching conditions to the semileptonic operators computed in~\cite{Fuentes-Martin:2019ign}. As an application of this result, we present a detailed discussion of the relative weight of vector and scalar contributions to the $b\to c\tau\nu$ decay amplitude. On the other hand, since vector-like fermions introduce a  new source of flavor violation with respect to the minimal version of the model, we present a systematic analysis of all the flavor-changing neutral-current (FCNC) amplitudes generated beyond the tree level at $\cO(\alpha_4)$. The latter effects turn out to be particularly relevant for processes such as $b \to s \nu\bar \nu$ or $B$--$\bar B$ mixing, which do not receive a tree-level contribution in this class of models. Combing the NLO amplitudes computed in this paper with those analyzed in~\cite{Fuentes-Martin:2019ign}, we present the first complete analysis of the $U_1$ impact in $b \to s \nu\bar \nu$ decays, which is of great phenomenological interest.

The structure of the paper is as follows: In Section~\ref{sec:model}, we introduce the minimal version of the model and the relevant interactions for the loop computations, and discuss in detail the effect of including vector-like fermions. In Section~\ref{sec:VLFCNC} we present our results of the loop-induced FCNCs. The phenomenological implications in $b \to s \nu\bar \nu$ and  $b\to c\tau\nu$ transitions are discussed in Section~\ref{sec:pheno}. The results are summarized in Section~\ref{sec:conclusions}. Appendices~\ref{app:VLmodels} and~\ref{app:loopDetails} provide further details on the vector-like fermion implementation and on the loop computations, respectively.

\section{The model}\label{sec:model}

\subsection{Minimal field content}

The 4321 models are based on the $SU(4)\times SU(3)^\prime\times SU(2)_L\times U(1)_X$ gauge symmetry. We denote the corresponding gauge fields by $H_\mu^A$, $C_\mu^a$, $W_\mu^I$ and $B^\prime_\mu$, with indices $A=1,\dots,15$, $a=1,\dots,8$ and $I=1,2,3$, and the gauge couplings by $g_4$, $g_3$, $g_2$ and $g_1$. The SM gauge group corresponds to the 4321 subgroup $SU(3)_c\times U(1)_Y\equiv[SU(4)\times SU(3)^\prime\times U(1)_X]_{\rm diag}$, with $SU(2)_L$ being the SM one. The hypercharge, $Y$, is defined in terms of the $U(1)_X$ charge, $X$, and the $SU(4)$ generator $T_4^{15}=\frac{1}{2\sqrt{6}}\mathrm{diag}(1,1,1,-3)$ by $Y=X+\sqrt{2/3}\,T_4^{15}$.

As in the SM case, it is useful to define the mixing angles $\theta_{1,3}$, relating the 4321 gauge couplings to the SM ones
\begin{align}
    g_s&=g_4\,s_3=g_3\, c_3\,, &
    g_Y&=\sqrt{\tfrac{3}{2}}\,g_4\, s_1=g_1\, c_1\,,
\end{align}
with $g_s$ and $g_Y$ denoting the $SU(3)_c$ and $U(1)_Y$ gauge couplings, and where we used a shorthand notation for the sine ($s_{1,3}$) and cosine ($c_{1,3}$) of the mixing angles. 

The SM gluon, $G_\mu^a$, and hypercharge gauge boson, $B_\mu$, written in terms of 4321 gauge bosons and mixing angles, read
\begin{align}
    G_\mu^a&=c_3\,C_\mu^a+s_3\,H_\mu^a\,,&
    B_\mu&=c_1\,B_\mu^\prime+s_1\,H_\mu^{15}\,.
\end{align}
The additional gauge bosons transform under the SM gauge group as $U_1\sim(\bf{3},\bf{1},2/3)$, $G^\prime\sim(\bf{8},\bf{1},0)$ and $Z^\prime\sim(\bf{1},\bf{1},0)$. In terms of the 4321 gauge eigenstates, they are given by
\begin{align}
    G_\mu^{\prime\,a}&=-s_3\,C_\mu^a+c_3\,H_\mu^a\,,\qquad
    Z_\mu^\prime=-s_1\,B_\mu^\prime+c_1\,H_\mu^{15}\,\no\\
    U_\mu^{1,2,3}&=\frac{1}{\sqrt{2}}(H_\mu^{9,11,13}-iH_\mu^{10,12,14})\,.
\end{align}
These gauge bosons become massive after the spontaneous breaking $SU(4) \times SU(3)^\prime\times U(1)_X \to SU(3)_c\times U(1)_Y$. The corresponding masses depend on the explicit form in which the 4321 model is spontaneously broken. In most 4321 models, this is triggered by the vacuum expectation values (vevs) of two scalar fields transforming in the antifundamental of $SU(4)$, $\Omega_1$ and  $\Omega_3$, singlet and triplet under $SU(3)^\prime$, respectively.\footnote{An additional scalar field, transforming in the adjoint of $SU(4)$ and singlet under the rest, is often introduced is some 4321 models~\cite{DiLuzio:2018zxy,Cornella:2019hct}. For simplicity, we only consider this field in Appendix~\ref{app:VLmodels}.} In this case, the gauge boson masses read
\begin{align}\label{eq:SU4vectorMasses}
m_G&=\frac{g_4}{2c_3}\,\sqrt{2\omega^2_3}\,,  \qquad 
\mZ=\frac{g_4}{2c_1}  \sqrt{\frac{3  \omega_1^2+  \omega_3^2}{2}}\,,\no\\
m_U&= \frac{g_4}{2} \sqrt{\omega_1^2+\omega_3^2}\,,
\end{align}
where $\omega_{1,3}$ are the $\Omega_{1,3}$ vacuum expectation values. In the limit $\omega_1=\omega_3$ and $g_{3,1}=0$ the massive vectors are degenerate: this is the result of an unbroken global symmetry, that we denote as $SU(4)_V$ custodial symmetry. The latter is defined by the diagonal combination of the $SU(4)\times SU(4)^\prime$ groups, with $SU(4)^\prime$ being the global group that contains $SU(3)^\prime$ and (part of) $U(1)_X$ as subgroup.

Electroweak symmetry breaking proceeds as in the SM through the vev of a SM-like Higgs, which could either be fundamental or composite~\cite{Fuentes-Martin:2020bnh}.

The minimal matter content of the model and their 4321 representations are described in Table~\ref{tab:minimal_content}. The $\Omega_{1,3}$ scalar fields decompose under the SM subgroup as
\begin{widetext}
\begin{align}\label{eq:Omega13}
\begin{aligned}
\Omega_1^\dagger&=\frac{\omega_1}{\sqrt{2}}
\begin{pmatrix}
\frac{g_4}{\sqrt{2}}\frac{1}{m_U}\big(\phi_U+\cot\beta\, h_U\big) \\[10pt]
1+\frac{S_1}{\omega_1}+i\,\frac{-3g_4}{2\sqrt{6}}\frac{1}{m_{Z^\prime}}\big(\phi_{Z^\prime}-\frac{1}{\sqrt{3}}\cot\beta\,h_{Z^\prime}\big)
\end{pmatrix}
\,,\\[5pt]
\Omega_3^\dagger&=\frac{\omega_3}{\sqrt{2}}
\begin{pmatrix}
\mathbb{1}_{3\times3}\,\big[1+\frac{1}{\sqrt{3}}\frac{S_3}{\omega_3}+i\,\frac{g_4}{2\sqrt{6}}\frac{c_1}{m_{Z^\prime}}\big(\phi_{Z^\prime}+\sqrt{3}\tan\beta\,h_{Z^\prime}\big)\big]+g_4\,T^a\,\frac{c_3}{m_{G^\prime}}\,\big(i\,\phi^a_{G^\prime}+h_{G^\prime}^a\big) \\[10pt]
-\frac{g_4}{\sqrt{2}}\frac{1}{m_U}\big(\phi_U^\dagger-\tan\beta\,h_U^\dagger\big) 
\end{pmatrix}
\,,
\end{aligned}
\end{align}
\end{widetext}
where $\phi_i$ and $h_i$ are, respectively, would-be Goldstone and physical scalars with the same quantum numbers as the corresponding gauge fields, $S_1$ and $S_3$ are SM singlet physical scalars, and $\tan\beta\equiv \omega_1/\omega_3$. In the limit of heavy radial modes ($m_{h_i},m_{S_{1,3}} \gg m_{U,Z^\prime,G^\prime}$), we are left with a non-linear realization of the $SU(4) \times SU(3)^\prime\times U(1)_X \to SU(3)_c\times U(1)_Y$ symmetry breaking, as in the composite model in~\cite{Fuentes-Martin:2020bnh}. As we show in Section~\ref{sec:VLFCNC}, most NLO corrections can be evaluated also in the non-linear case with marginal ambiguities on the size of the effects.

\begin{table}[t]
\centering
\setlength{\tabcolsep}{5pt}
\renewcommand{\arraystretch}{1.1}
\begin{tabular}{|c|c|c|c|c|c|}
\hline
Field & $SU(4)$ & $SU(3)^\prime$ &  $SU(2)_L$  & $U(1)_X$ \\
\hline
\hline 
$\psi_L$ & $\mathbf{4}$  & $\mathbf{1}$  & $\mathbf{2}$  &  0 \\
$\psi^+_R$ & $\mathbf{4}$ & $\mathbf{1}$   & $\mathbf{1}$  &  $1/2$   \\ 
$\psi^-_R$ & $\mathbf{4}$ & $\mathbf{1}$   & $\mathbf{1}$  &  $-1/2$   \\ 
$q_L^{\prime\,i}$ & $\mathbf{1}$  & $\mathbf{3}$ & $\mathbf{2}$  &  $1/6$   \\[2pt]
$u_R^i$ & $\mathbf{1}$  & $\mathbf{3}$ & $\mathbf{1}$  &  $2/3$   \\[2pt]
$d_R^i$ & $\mathbf{1}$  & $\mathbf{3}$ & $\mathbf{1}$  &  $-1/3$   \\[2pt]
$\ell_L^{\prime\,i}$ & $\mathbf{1}$  & $\mathbf{1}$ & $\mathbf{2}$  &  $-1/2$   \\[2pt]
$e_R^i$ & $\mathbf{1}$  & $\mathbf{1}$ & $\mathbf{1}$  &  $-1$   \\
\hline
\hline
$H$ & $\mathbf{1}$ & $\mathbf{1}$ & $\mathbf{2}$ & $1/2$\\
$\Omega_3$ & $\mathbf{\bar 4}$ & $\mathbf{3}$ & $\mathbf{0}$ & $1/6$\\
$\Omega_1$ & $\mathbf{\bar 4}$ & $\mathbf{1}$ & $\mathbf{0}$ & $-1/2$\\
\hline
\end{tabular}
\caption{Minimal matter content. Here $i=1,2$, $\psi_L\equiv(q_L^{\prime 3}\; \ell_L^{\prime 3})^\intercal$, $\psi_R^+\equiv(u_R^3\; \nu_R^ 3)^\intercal$ and $\psi_R^-\equiv(d_R^3\; e_R^3)^\intercal$. The prime in the left-handed fields indicate that these are not mass eigenstates (see~\ref{sec:VLferm}).}
\label{tab:minimal_content}
\end{table}

We consider a version of the 4321 model were the would-be SM fields (in the absence of fermion mixing) are charged non-universally under the 4321 gauge group, see Table~\ref{tab:minimal_content}. The fermion content charged under $SU(4)$ consists of three fields transforming as Pati-Salam representations under $SU(4)\times SU(2)_L\times U(1)_X$: one $SU(2)_L$ doublet, $\psi_L$, and two $SU(2)_L$  singlets, $\psi_R^\pm$. In addition, we have two identical SM-like families, singlets under $SU(4)$ and transforming as the SM fermions under $SU(3)^\prime\times SU(2)_L \times U(1)$. In the absence of fermion mixing (see~\ref{sec:VLferm}), the $SU(4)$-charged fermions would correspond to the SM third generation (plus a right-handed neutrino), and the $SU(4)$-singlets to the light-generation SM fermions.

\subsection{Relevant interactions}
We describe only those interactions that are relevant for the loop computations below. The $U_1$ interactions with SM gauge bosons are given by
\begin{align}\label{eq:LQLag}
\cL &\supset-\frac{1}{2}\,U_{\mu\nu}^\dagger\, U^{\mu\nu} -ig_s\,U_\mu^\dagger\,T^a\,U_\nu\,G^{a\,\mu\nu} \no\\
&\quad-\frac{2}{3}\,ig_Y \,U_\mu^\dagger\,\,U_\nu\,B^{\mu\nu}\,,
\end{align}
where $U_{\mu \nu} = D_\mu U_\nu-D_\nu U_\mu$, with $D_\mu = \partial_\mu - i g_s\, G_\mu^aT^a - i \frac{2}{3} g_Y  B_\mu$. If we neglect terms of $\mathcal{O}(g_{\rm SM}^2)$, with $g_{\rm SM}$ being any of the SM couplings, the triple gauge interactions of two $U_1$ with $Z^\prime$ ($G^\prime$) are the same as with $B$ ($G$) with the replacement $g_Y\to g_4\sqrt{3/2}$ ($g_s\to g_4$). The relevant interactions of Goldstone and radials to gauge bosons read
\begin{align}
\mathcal{L}&\supset \frac{g_4}{2\sqrt{6}}\,Z^\prime_\mu\big[
(3-2\sin^2\beta)\,\phi_U^\dagger\,i\partial^\mu \phi_U\no\\[-5pt]
&\hspace{1.8cm}+(1+2\sin^2\beta)\,h_U^\dagger\,i\partial^\mu h_U\no\\
&\hspace{1.8cm}-2\sin\beta\cos\beta\;(h_U^\dagger\,i\partial^\mu\phi_U+\phi_U^\dagger\,i\partial^\mu h_U) \no\\
&\hspace{1.8cm}+2m_U\, (1-2\sin^2\beta)\,\phi_U^\dagger\,U^\mu\no\\
&\hspace{1.8cm}-4m_U \sin\beta\cos\beta\,h_U^\dagger\,U^\mu+{\rm h.c.}\big] \no\\
&\quad +g_4\,G_\mu^{\prime\,a}\,\big[\sin^2\beta\,\phi_U^\dagger\,T^a\,i\partial^\mu \phi_U\no\\
&\hspace{1.8cm}+\cos^2\beta\,h_U^\dagger\,T^a\,i\partial^\mu\,h_U \no\\
&\hspace{1.8cm}+\sin\beta\cos\beta\;(h_U^\dagger\,T^a i\partial^\mu\,\phi_U+\phi_U^\dagger\,T^a\,i\partial^\mu h_U) \no\\
&\hspace{1.8cm}-m_U\,(1-2 \sin^2\beta)\,\phi_U^\dagger\,T^a\,U^\mu\no\\
&\hspace{1.8cm}+2 m_U \sin\beta\cos\beta\,h_U^\dagger\,T^a\,U^\mu+{\rm h.c.}\big]\,,
\end{align}
with $T^a$ being the $SU(3)$ generators. In the absence of fermion mixing (see section below), and neglecting once more terms of $\mathcal{O}(g_{\rm SM}^2)$, the interactions between the massive vectors and fermions read
\begin{align}
\cL &\supset    \frac{g_4}{\sqrt{2}}\, U_\mu\,(\bar q_L^{\prime 3}\,\gamma^\mu\ell_L^{\prime 3}+\bar q_R^3\,\gamma^\mu\ell_R^3 ) +{\rm h.c.} \no\\
&\quad + g_4\, \Zp_\mu \,\big(\bar\psi_L\,\gamma^\mu\,T^{15}\psi_L+\bar\psi_R\,\gamma^\mu\,T^{15}\psi_R\big)\no\\
& \quad +  g_4\, \Gp_\mu\,\big(  \bar q_L^{\prime 3}\gamma^\mu\,T^a\bar q_L^{\prime 3}+ \bar q_R^ 3\gamma^\mu\,T^a\bar q_R^3\big)\,,  
\label{eq:U1Lag}
\end{align}
where $\psi_R=(\psi_R^+\;\psi_R^-)^\intercal$, $q_R^3=(u_R^3\;d_R^3)^\intercal$, $\ell_R^3=(\nu_R^3\;e_R^3)^\intercal$. Finally, the couplings of Goldstones and radials to fermions depend on the specific vector-like implementation (see section below) and are described in Appendix~\ref{app:VLmodels}.

\begin{table}[t]
\begin{center}
\setlength{\tabcolsep}{5pt}
\renewcommand{\arraystretch}{1.1}
\begin{tabular}{|c|c|c|c|c|c|c|}
\hline
Model & Field & $SU(4)$ & $SU(3)^\prime$ &  $SU(2)_L$  & $U(1)_X$ \\
\hline
& $\chi_L$ & $\mathbf{4}$  & $\mathbf{1}$  & $\mathbf{2}$  &  0 \\ 
I & $Q_R$ & $\mathbf{1}$  & $\mathbf{3}$  & $\mathbf{2}$  &  $1/6$ \\ 
& $L_R$ & $\mathbf{1}$  & $\mathbf{1}$ &  $\mathbf{2}$  & $-1/2$ \\
\hline 
\hline 
\multirow{2}{*}{II}& $\chi_L$ & $\mathbf{4}$  & $\mathbf{1}$  & $\mathbf{2}$  &  0 \\ 
& $\chi_{R}$ & $\mathbf{4}$  & $\mathbf{1}$ &  $\mathbf{2}$  & 0 \\
\hline
\end{tabular}
\end{center}
\caption{Additional fermion content. Here $\chi_L=(Q_L^\prime\;L_L^\prime)^\intercal$ and $\chi_R=(Q_R\;L_R)^\intercal$. The prime in the $\chi_L$ components indicates that these are not mass eigenstates.}
\label{tab:VLcontent}
\end{table}

\subsection{Vector-like fermions}\label{sec:VLferm}

We now discuss the inclusion of massive fermions, vector-like under the SM gauge group, to the minimal model discussed in the previous section. In realistic 4321 models, these are introduced to induce couplings between the $SU(4)$ vectors and the light SM families. For simplicity, here we focus on the mixing with a single SM-like family, and therefore introduce only one vector-like family. More precisely, we add to the minimal model one family of left-handed fermions, transforming in the fundamental representations of $SU(4)$ and $SU(2)_L$, and one family of right-handed partners. The massive fermions are vector-like under the SM gauge group, therefore, the right-handed partners should transform in the fundamental of $SU(2)_L$, but there is freedom in the $SU(4)\times SU(4)^\prime$ transformations.  As shown in Table~\ref{tab:VLcontent}, we consider two possible implementations: They are either $SU(4)$ singlets and transform in the fundamental of $SU(4)^\prime$ (model I), as in~\cite{Fuentes-Martin:2020bnh}; or they are $SU(4)^\prime$ singlets and transform in the fundamental of $SU(4)$ (model II), as in~\cite{Cornella:2019hct}. 

Having two $SU(4)$ charged fields with the same Lorentz and gauge transformation properties, $\psi_L$ and $\chi_L$, leads to a new flavor symmetry that we denote $U(2)_\xi$,
where 
\begin{align}
\xi_L = 
\begin{pmatrix}
\psi_L \\
\chi_L
\end{pmatrix}
\,.
\label{eq:xidef}
\end{align}
This symmetry is broken by the fermion masses, giving rise to a possible mixing among $\psi_L$ and $\chi_L$, and possibly also the $SU(4)$-singlet fermions, $q_L^\prime$ and $\ell_L^\prime$, after the breaking of the $SU(4)$ symmetry. The $SU(4)$ breaking in the fermion masses could either be due to the vevs of $\Omega_{1,3}$ or via new sources. We discuss the details for each implementation in Appendix~\ref{app:VLmodels}.  

In either case, the mass terms after $SU(4)$ breaking read
\begin{align}\label{eq:MassMix}
\cL_{\rm mass}  = 
\bar\Psi_L^{q\,\prime}\, M_q\,Q_R+\bar\Psi_L^{\ell\,\prime}\, M_\ell\,L_R\,,
\end{align}
with the left-handed fermions arranged into the the flavor vectors
\begin{align}\label{eq:Psiprime}
\Psi^{q\,\prime}_L&=
\begin{pmatrix}
q_L^{\prime\, 2}\\[2pt]
q_L^{\prime\, 3}\\[2pt]
Q_L^\prime
\end{pmatrix}
\,,&
\Psi^{\ell\,\prime}_L&=
\begin{pmatrix}
\ell_L^{\prime\, 2}\\[2pt]
\ell_L^{\prime\, 3}\\[2pt]
L_L^\prime
\end{pmatrix}
\,.
\end{align}
and where $M_{q,\ell}$ are $3$-dimensional mass vectors. Without loss of generality, these mass vectors can be written as 
\begin{align}
M_q&=\tilde W_q\,O_q\begin{pmatrix} 0 & 0 & m_Q \end{pmatrix}^\intercal\,,\no\\
M_\ell&=\tilde W_\ell\,O_\ell\begin{pmatrix} 0 & 0 & m_L \end{pmatrix}^\intercal\,,
\label{eq:M99}
\end{align}
where $m_{Q,L}$ are the vector-like fermion masses. Here, the $3\times3$ orthogonal matrices $O_{q,\ell}$ parametrize the mixing among different $SU(4)$ representations, and take the explicit form
\begin{align}\label{eq:Omatrices}
O_{q,\ell}=
\begin{pmatrix}
c_{Q,L} & 0 & s_{Q,L}\\
0 & 1 & 0\\
-s_{Q,L} & 0 & c_{Q,L}\\
\end{pmatrix}
\,,
\end{align}
with $s_{Q,L}\, (c_{Q,L})$ being the sine (cosine) of the $\theta_{Q,L}$ mixing angles.

On the other hand, the $3\times3$ unitary matrices $\tilde W_{q,\ell}$ parametrize the mixing among $SU(4)$ states, and can be decomposed as
\begin{align}
    \tilde W_{q,\ell}=
    \begin{pmatrix}
    1 & 0\\
    0 & W_{q,\ell}
    \end{pmatrix}
    \,,
\end{align}
with $W_{q,\ell}$ being unitary $2\times 2$ matrices. To better understand the origin of the flavor mixing matrices, it is convenient to decompose the mass vector in~\eqref{eq:MassMix} into two components
\begin{align}\label{eq:VLMass14}
    M_q&=
    \begin{pmatrix}
    M_q^1\\[2pt]
    M_q^4
    \end{pmatrix}
    \,,&
    M_\ell&=
    \begin{pmatrix}
    M_\ell^1\\[2pt]
    M_\ell^4
    \end{pmatrix}
    \,.
\end{align}
where $M_{q,\ell}^1$ is real and $M_{q,\ell}^4$ is a 2-vector. Their combined presence encodes two different flavor symmetry breakings:
\begin{itemize}
    \item[i)] The $U(2)_\xi$ alignment of $M_q^4$ and $M_\ell^4$ is at the origin of the $W_{q,\ell}$ matrices. Indeed, we have
    \begin{align}
        M_q^4&=W_q\,(0\;M_3)^\intercal\,,\nonumber\\
        M_\ell^4&=W_\ell\,(0\;M_1)^\intercal\,,
    \end{align}
    with $M_{3,1}$ being real parameters with mass dimension, and $W_{q,\ell}$ as before. The $SU(4)$ breaking from $M_{q,\ell}^4$ is analogous to the $SU(2)_L$ breaking in the SM from the up- and down-type fermion masses. As we show below, only the misalignment of quarks and leptons in $U(2)_\xi$ space, encoded in $W=W_q^\dagger\, W_\ell$,  is physical.

    \item[ii)] The ratio between $M_{q,\ell}^1$ and $M_{1,3}$ determines the breaking of the $U(2)_{q,\ell}$ flavor symmetry of the light fermions. Such breaking appears in the form of the $O_{q,\ell}$ mixing matrices, with $\tan\theta_{Q,L}=M_{q,\ell}^1/M_{3,1}$.
\end{itemize}

In the mass basis, the $SU(4)$ vector interactions with left-handed fermions (in the limit $g_{3,1}=0$) take the form
\begin{align}\label{eq:intMassBasis}
\mathcal{L}_{\rm int}&\supset\frac{g_4}{\sqrt{2}}\,U_\mu\,\bar \Psi^q_L\,O_q^\intercal\, P_{23}\, \tilde W\, O_\ell\,\gamma^\mu\Psi^\ell_L+{\rm h.c.} \no\\
&+\frac{g_4}{2\sqrt{6}}\,Z^\prime_\mu\,\left(\bar\Psi^q_L\,O_q^\intercal\, P_{23}\, O_L^q\,\gamma^\mu\,\Psi^q_L-3\,\bar\Psi^\ell_L\,O_\ell^\intercal\, P_{23}\, O_\ell\,\gamma^\mu\,\Psi^\ell_L\right)\no \\
&+g_4\,G^{\prime\,a}_\mu\,\bar\Psi^q_L\,O_q^\intercal\, P_{23}\, O_q\,\gamma^\mu\,T^a\,\Psi^q_L\,,
\end{align}
with $P_{23}\equiv\mathrm{diag}(0,1,1)$ projecting into the $SU(4)$ components of $\Psi_L^{q,\ell}$. Since $\tilde W_{q,\ell}$ and $P_{23}$ commute, the individual $\tilde W_{q,\ell}$ matrices are not observable, but only the combination 
\begin{align}\label{eq:Wtilde}
\tilde W=\tilde W_q^\dagger\, \tilde W_\ell=
\begin{pmatrix}
1 & 0\\
0 & W\\
\end{pmatrix}
\,.
\end{align}
It is convenient to rewrite these interactions in an $SU(4)$ basis, or in the quark ($\mathcal{Q}_L^i$) and lepton ($\mathcal{L}_L^i$) components of $\xi^i_L$, that in the mass-eigenstate
basis are given by
\begin{align}\label{eq:SU4basis}
\begin{pmatrix}
0\\[2pt]
\mathcal{Q}_L^1\\[2pt]
\mathcal{Q}_L^2
\end{pmatrix}
=
P_{23}\,O_q
\begin{pmatrix}
q_L^2\\[2pt]
q_L^3\\[2pt]
Q_L
\end{pmatrix}
\,,\quad
\begin{pmatrix}
0\\[2pt]
\mathcal{L}_L^1\\[2pt]
\mathcal{L}_L^2
\end{pmatrix}
=
P_{23}\,O_\ell
\begin{pmatrix}
\ell_L^2\\[2pt]
\ell_L^3\\[2pt]
L_L
\end{pmatrix}
\,.
\end{align}
In this basis, the interactions in~\eqref{eq:intMassBasis} take the simple form ($i=1,2$)
\begin{align}
\begin{aligned}
\mathcal{L}&\supset\frac{g_4}{\sqrt{2}}\,U_\mu\,\mathcal{\bar Q}_L^i\,W_{ij}\,\gamma^\mu\, \mathcal{L}_L^j+{\rm h.c.}\\
&+\frac{g_4}{2\sqrt{6}}\,Z^\prime_\mu\left(\mathcal{\bar Q}_L^i\gamma^\mu \mathcal{Q}_L^i-3\,\mathcal{\bar L}_L^i\gamma^\mu \mathcal{L}_L^i\right)\\
&+g_4\,G^{\prime\,a}_\mu\,\mathcal{\bar Q}_L^i\gamma^\mu\, T^a \mathcal{Q}_L^i\,,
\end{aligned}
\end{align}
The unitary matrix $W$ can be regarded as a generalization of the CKM matrix to $SU(4)$ or quark-lepton space. Similarly to the CKM case, the $W$ matrix is the only source of flavor-changing transitions among $SU(4)$ states, and it appears only in interactions involving both quarks and leptons. In this sense, the vector LQ, $U_\mu$, is analogous to the SM $W_\mu$. Similarly, the $Z^\prime_\mu, G^\prime_\mu$ are analogous to the SM $Z_\mu$ and their interactions are $SU(4)$ flavor-conserving at tree-level. In analogy to the SM, we will denote $U_\mu$ transitions as charged current and $Z^\prime_\mu, G^\prime_\mu$ transitions as neutral currents. As in the SM, flavor-changing neutral currents (FCNCs) proportional to the $W$ matrix are generated at the loop level. We compute these contributions in Section~\ref{sec:VLFCNC}. 

Finally, note that the structure in \eqref{eq:MassMix} holds in the limit of unbroken $SU(2)_L$ symmetry, with a single family of $SU(4)$-singlet fermions (corresponding to the SM with 2 generations). Its generalization to a 3 generation case, and the inclusion of $SU(2)_L$-breaking effects from the SM Yukawa couplings is straightforward, as long as 
we neglect light-quark mass effects. Note in particular that the 2-3 mixing form the SM Yukawa couplings (corresponding to a 1-2 mixing in the $\Psi_L^{q,\ell}$ space) can effectively be encoded via the replacement $O_{q,\ell} \to O_{q,\ell}\, L_{q,\ell}$, where $L_{q,\ell}$ are rotation matrices in 1-2 space resulting from the diagonalization of the SM Yukawa couplings.

\section{FCNC four-fermion and dipole operators}\label{sec:VLFCNC}

\subsection{Generalities}\label{subsec:FCNCgen}
Before presenting the results, it is illustrative to show explicitly the unitarity cancellations taking place in the FCNC loops, analogous to the so-called Glashow--Iliopoulos--Maiani (GIM) mechanism in the SM. For instance, for the fermion self-energy graph shown in diagram (i) in Figure~\ref{fig:legs} we have
\begin{widetext}
\begin{align}
\begin{aligned}
\mathcal{A}_{\rm self}&\propto\bar\Psi_q\,O_q^\intercal\, P_{23}\,\tilde W\,O_\ell\left[\mathbb{1}\,f_2^\psi(s,m_U,0)+P_3\,\big[f_2^\psi(s,m_U,m_Q)-f_2^\psi(s,m_U,0)\big]\right]\,O_\ell^\intercal\,\tilde W^\dagger\,P_{23}\,O_q\,\Psi_q \\
&=\bar\Psi_q\,O_q^\intercal\,P_{23}\,\left[\mathbb{1}\,f_2^\psi(s,m_U,0)+c_L^2\,\tilde W\,P_3\,\tilde W^\dagger\big[f_2^\psi(s,m_U,m_Q)-f_2^\psi(s,m_U,0)\big] \right]P_{23}
\,O_q\,\Psi_q\\
&=\mathcal{\bar Q}_L\,\left[\mathbb{1}\,f_2^\psi(s,m_U,0)+c_L^2\,\tilde W\,P_3\,\tilde W^\dagger\big[f_2^\psi(s,m_U,m_Q)-f_2^\psi(s,m_U,0)\big] \right]\mathcal{Q}_L\,,
\end{aligned}
\end{align}
\end{widetext}
where $f_2^\psi$ is the loop function, we took $\ell_L^i$ massless, and $P_3=\mathrm{diag}(0,0,1)$. In the second line, we used the property $P_{23}\,\tilde W=\tilde W\,P_{23}$, $\tilde W$ ($O_\ell$) unitarity (orthogonality), and the following relation
\begin{align}
P_{23}\;O_\ell\,P_3\,O_\ell^\intercal\,P_{23}=c_L^2\,P_3\,.
\end{align}
Similar unitarity cancellations also take place in vertices and boxes. It is worth stressing some features that are common to all the FCNC loops presented here: 
\begin{itemize} 
\item[i.] Since we are dealing with $SU(4)$ interactions only, the external states can always be written in the $SU(4)$ basis defined in~\eqref{eq:SU4basis}.
\item[ii.]  Similarly to the SM, the $SU(4)$ flavor-changing contribution is proportional to $\tilde W\,P_3\,\tilde W^\dagger=W_{i2}\,W_{j2}^*$. The effect of $O_\ell$ is seen in the factor $c_L^2$, which gives the projection of the massive component in the $SU(4)$ state.
\item[iii.] The FCNC part of the amplitude is proportional to the flavor- and $SU(4)_V$-breaking component of the vector-like mass in \eqref{eq:MassMix}:
in the limit of small breaking  $c_L^2 W_{i2}\,W_{j2}^* \approx W_{i\not=j}$ and, by means of Eq.~(\ref{eq:Deltapert}), we can interpret the flavor-violating amplitude as the result of inserting the symmetry-breaking mass term on the vector-like fermion propagator. 

\item[iv.] While our computations present many similarities with those in the SM, one should not be tempted to simply rescale the SM contributions. Indeed, the presence of both $W$ and $\mathcal{O}_{q,\ell}$ mixing matrices, instead of just the CKM matrix, yield loop functions that are different from their SM analog. In the limit of small breaking, this can be understood from the fact that symmetry breaking terms and fermion masses (controlling the loop functions) can be varied independently in our case, while they are in one-to-one correspondence in the SM. 
\end{itemize}

In the next subsection, we present the result of the effective flavor-changing vertices of the $Z^\prime$ and $G^\prime$ massive vectors to fermions, using the $SU(4)$ basis in~\eqref{eq:SU4basis}. These (gauge dependent) vertices, which are evaluated in the Feynman gauge, are then combined with the box amplitudes in order to obtain the (gauge independent) contributions to the Wilson coefficients (WC) of the semileptonic FCNC operators (written in the SMEFT basis). In subsection~\ref{subsec:dipoles}, we present the results of dipole-type effective operators, and in subsection~\ref{subsec:DF2} of the $\Delta F=2$ hadronic operators. 

\vspace{1.7cm}

\subsection{\texorpdfstring{$Z^\prime$}{Zp} and \texorpdfstring{$G^\prime$}{Gp} flavor-changing vertices}\label{subsec:FCNCVert}

We define the following effective vertices that encode the FCNCs
\begin{align}
    \mathcal{L}&\supset\frac{g_4}{2\sqrt{6}}\,Z_\mu^\prime\big[(\mathcal{\bar Q}_L^i\gamma^\mu \,\Gamma_{Z^\prime_q}^{ij}\mathcal{Q}_L^j)-3\,(\mathcal{\bar L}_L^\alpha\gamma^\mu \,\Gamma_{Z^\prime_\ell}^{\alpha\beta}\,\mathcal{L}_L^\beta)\big]\no\\
    &\quad+g_4\,G_\mu^{\prime\,a}(\mathcal{\bar Q}_L^i\gamma^\mu \,T^a\,\Gamma_{G^\prime}^{ij}\mathcal{Q}_L^j)\,,
\end{align}
where
\begin{align}\label{eq:effectiveVert}
\Gamma_{Z^\prime_q}^{ij}&=\frac{\alpha_4}{8\pi}\,W_{i2}W_{j2}^*\,c_L^2\,V_{Z^\prime_q}(x_{Z^\prime},x_L,\theta_L)\,,\no\\
\Gamma_{Z^\prime_\ell}^{\alpha\beta}&=\frac{\alpha_4}{8\pi}\,W_{2\alpha}^*W_{2\beta}\,c_Q^2\,V_{Z^\prime_\ell}(x_{Z^\prime},x_Q,\theta_Q)\,,\no\\
\Gamma_{G^\prime}^{ij}&=\frac{\alpha_4}{8\pi}\,W_{i2}W_{j2}^*\,c_L^2\,V_{G^\prime}(x_{Z^\prime},x_L,\theta_L)\,.
\end{align}
where $x_{Z^\prime, G^\prime}=m_{Z^\prime, G^\prime}^2/m_U^2$ and $x_{Q,L}=m_{Q,L}/m_U$. The gauge and Goldstone contributions to the vertex functions $V_\mathcal{V}(x_\mathcal{V},x_i,\theta_i)$, with $\mathcal{V}=Z_q^\prime,Z_\ell^\prime,G^\prime$, have the general form 
\begin{align}\label{eq:Vfunc}
    V_\mathcal{V}^M(x_\mathcal{V},x_i,\theta)=x_i\!\left[d_\mathcal{V}^M(x_\mathcal{V})\,\Delta_U+F_\mathcal{V}^M(x_\mathcal{V},x_i,\theta)\right],
\end{align}
with $\Delta_U=\frac{1}{\epsilon}-\gamma_E+\ln 4\pi+\ln \frac{\mu^2}{m_U^2}$, and $M=I,II$ denotes the different vector-like models in Table~\ref{tab:VLcontent}. As expected, the unitarity cancellation discussed in the previous section ensures that the flavor-changing vertices vanish in the limit $x_i\to0$. The loop functions $F_\mathcal{V}$ are given in Appendix~\ref{app:FCNCvertFinal}. For reference, we give the value of $F_\mathcal{V}$ in the limit $x_\mathcal{V}=x_i=1$ and neglecting terms of $\mathcal{O}(\theta_i^2)$:
\begin{align}
F_{Z^\prime_q}^I(1,1,\theta_L)&=-\frac{5}{2}\,, &
F_{Z^\prime_q}^{II}(1,1,\theta_L)&=-4\,,\no\\
F_{Z^\prime_\ell}^I(1,1,\theta_Q)&=-\frac{7}{2}\,, &
F_{Z^\prime_\ell}^{II}(1,1,\theta_Q)&=-4\,,\no\\
F_{G^\prime}^I(1,1,\theta_L)&=-1\,, &
F_{G^\prime}^{II}(1,1,\theta_L)&=-1\,.
\end{align}
The functions multiplying the divergent piece read \begin{align}\label{eq:Div} 
d_{Z^\prime_q}^I(x_{Z^\prime})&=\frac{3}{2}-x_{Z^\prime}\,, &
d_{Z^\prime_q}^{II}(x_{Z^\prime})&=-x_{Z^\prime}\,,\no\\
d_{Z^\prime_\ell}^I(x_{Z^\prime})&=\frac{1}{2}-x_{Z^\prime}\,, &
d_{Z^\prime_q}^{II}(x_{Z^\prime})&=-x_{Z^\prime}\,,\no\\
d_{G^\prime}^I(x_{G^\prime})&=-\frac{1}{4}\,x_{G^\prime}\,, &
d_{G^\prime}^{II}(x_{G^\prime})&=-\frac{1}{4}\,x_{G^\prime}\,.
\end{align}
In a renormalizable model, we expect $d_\mathcal{V}=0$, since the FCNC vertices are not present at tree level. Indeed, this is the case also in our models, but only after the introduction of the radial contributions. These depend on the different implementations of the scalar sector, and thus should be discussed for each model separately.

\subsubsection{Model I}

The scalar content of this model is the same as the one described in Section~\ref{sec:model}. The only radial mode that can mediate flavor-changing transitions proportional to the $W$ matrix is the LQ radial $h_U$ (see~\eqref{eq:Omega13}). Similarly to what we did with the gauge and Goldstone contributions, we decompose the contribution from the scalar LQ as
\begin{align}
V_\mathcal{V}^R(x_\mathcal{V},x_R,\tilde x_i,x_i)&=x_i\,d_\mathcal{V}^I(x_\mathcal{V})\no\\
&\quad\times\left[-\Delta_U+F_\mathcal{V}^R(x_\mathcal{V},x_R,\tilde x_i,x_i)\right].
\end{align}
with $x_R=m_{h_U}^2/m_U^2$ and $\tilde x_i=m_i^2/m_{h_U}^2$. As expected, the LQ radial contribution cancel exactly the divergence from the Goldstone sector. The corresponding expressions for the $F_\mathcal{V}^R(x_\mathcal{V},x_R,\tilde x,x)$ loop functions are given in Appendix~\ref{app:FCNCvertFinal}. In the limit of heavy radials, i.e. $\tilde x_i\to0$ and $x_R\to\infty$, these reduce to
\begin{align}
F_\mathcal{V}^R(x_\mathcal{V},x_R,\tilde x_i,x)&\to\ln x_R-\frac{5}{2}\,.
\end{align}
Therefore, the net effect of the LQ radials in the heavy radial limit is to replace the divergence in \eqref{eq:Vfunc} by
\begin{align}
\Delta_U\to \ln x_R-\frac{5}{2}\,,
\end{align}
namely by a logarithm of the mass ratio plus an $\mathcal{O}(1)$ constant that is the same for all three effective vertices.

\subsubsection{Model II}

Apart from the $\Omega_{1,3}$ fields introduced in Section~\ref{sec:model}, this model requires an additional scalar with non-zero vev, $\Omega_{15}$, to generate a non-trivial $W$ matrix (see Appendix~\ref{app:VLmodels} for details). This field transforms in the adjoint of $SU(4)$ and therefore it contains a scalar LQ. In the limit $\omega_{1,3}=0$ (or equivalently $x_{Z^\prime,G^\prime}=0$), this LQ is identified with the would-be Goldstone boson and the gauge and Goldstone contributions become finite (see~\eqref{eq:Div}). However, in the general case where $\omega_{1,3}\neq0$, as needed to have $\theta_{L,Q}\neq0$ (see Appendix~\ref{app:VLmodels}), the Goldstone contribution is divergent and all LQ radials have to be considered. Since the scalar sector is more involved in this case, we do not compute the LQ radial contributions here. However, we note that, as in the case above, their effect in the heavy and degenerate radial mass limit is to replace the divergence in~\eqref{eq:Vfunc} by
\begin{align}
\Delta_U\to \ln x_R+f_R\,,
\end{align}
where, similarly to the case above, $x_R=m_R^2/m_U^2$, with $m_R$ being the mass of the LQ radials, and $f_R$ is a (universal) constant, expected to be of $\mathcal{O}(1)$. 

\subsection{\texorpdfstring{$\Delta F=1$}{DF=1} semileptonic operators}

We define the following effective Lagrangian for the semileptonic operators involving $\Delta F=1$ flavor-changing transitions that are absent at tree level 
\begin{align}
    \mathcal{L}_{\Delta Q=1}^{4F}&=-\frac{4G_U}{\sqrt{2}}\frac{\alpha_4}{4\pi}\,W_{12}^*W_{22}\,s_Q\, c_L^2\no\\
    &\quad\times[\mathcal{C}_{\ell q}^{\alpha\alpha23}\,\mathcal{O}_{\ell q}^{\alpha\alpha23}+\mathcal{C}_{qe}^{2333}\,\mathcal{O}_{qe}^{2333}]+{\rm h.c.}\,,\no\\
    \mathcal{L}_{\Delta L=1}^{4F}&=-\frac{4G_U}{\sqrt{2}}\frac{\alpha_4}{4\pi}\,W_{21}W_{22}^*\,s_L\, c_Q^2\no\\
    &\quad\times[\mathcal{C}_{\ell q}^{23ii}\,\mathcal{O}_{\ell q}^{23ii}+\mathcal{C}_{\ell u(d)}^{2333}\,\mathcal{O}_{\ell u(d)}^{2333}]+{\rm h.c}\,,
\end{align}
with $G_U=\sqrt{2}\,g_4^2/8 m_U^2$ and the effective operators
\begin{align}
\mathcal{O}_{\ell q}^{\alpha\beta ij}&=(\bar \ell_L^\alpha\gamma_\mu \ell_L^\beta)(\bar q_L^i\gamma^\mu q_L^j)\,,\no\\
\mathcal{O}_{qe}^{ij\alpha\beta}&=(\bar q_L^i\gamma_\mu q_L^j)(\bar e_R^\alpha\gamma^\mu e_R^\beta)\,,\no\\
\mathcal{O}_{\ell u}^{\alpha\beta ij}&=(\bar \ell_L^\alpha\gamma_\mu \ell_L^\beta)(\bar u_R^i\gamma^\mu u_R^j)\,,\no\\
\mathcal{O}_{\ell d}^{\alpha\beta ij}&=(\bar \ell_L^\alpha\gamma_\mu \ell_L^\beta)(\bar d_R^i\gamma^\mu d_R^j)\,.
\end{align}
Neglecting contributions of $\mathcal{O}(g_{\rm SM}^2/g_4^2)$, with $g_{\rm SM}$ being the SM couplings, the corresponding Wilson coefficients  at the matching scale are given by ($M=I,II$)
\begin{align}
\mathcal{C}_{\ell q}^{\alpha\alpha23}&=\delta_{\alpha3}\left[\frac{B_{q\ell}^{1211}}{W_{12}^*W_{22} \, c_L^2}+\frac{V_{Z_q}^M}{8x_{Z^\prime}}\right]\no\\
&\quad+\delta_{\alpha2}\,s_L^2\left[\frac{B_{q\ell}^{1222}}{W_{12}^*W_{22}\, c_L^2}+\frac{V_{Z_q}^M}{8x_{Z^\prime}}\right]\,,\no\\
\mathcal{C}_{qe}^{2333}&=\frac{B_{qe}^{1211}}{W_{12}^*W_{22}\, c_L^2}+\frac{V_{Z_q}^M}{8x_{Z^\prime}}\,,\no\\
\mathcal{C}_{\ell q}^{23ii}&=\delta_{i3}\left[\frac{B_{q\ell}^{1121}}{W_{21}W_{22}^*\, c_Q^2}+\frac{V_{Z_\ell}^M}{8x_{Z^\prime}}\right]\no\\
&\quad+\delta_{i2}\,s_Q^2\left[\frac{B_{q\ell}^{2221}}{W_{21}W_{22}^*\, c_Q^2}+\frac{V_{Z_\ell}^M}{8x_{Z^\prime}}\right]\,,\no\\
\mathcal{C}_{\ell u(d)}^{2333}&=\frac{B_{u(d)\ell}^{1121}}{W_{21}W_{22}^*\, c_Q^2}+\frac{V_{Z_\ell}^M}{8x_{Z^\prime}},
\end{align}
where we omitted the arguments in the loop functions to simplify the notation. The expressions for the $B^{ijkl}_{fg}$ loop functions, whose flavor indices refer to the $SU(4)$ basis, are given in Appendix~\ref{app:FCNCBoxes}. If we neglect terms of $\mathcal{O}(|W_{12}|^2)$, we find the following expression for the box functions
\begin{align}
\frac{B_{q\ell(e)}^{1211}}{W_{12}^*W_{22} \, c_L^2}&\approx 2 x_L \left(\frac{1}{1 - x_L} + \frac{\ln x_L}{(x_L - 1)^2}\right)\,,\no\\
\frac{B_{q\ell(e)}^{1121}}{W_{21}W_{22}^*\, c_Q^2}&\approx 2 x_Q \left(\frac{1}{1 - x_Q} + \frac{\ln x_Q}{(x_Q - 1)^2}\right)\,.
\end{align}

We also provide the corresponding amplitudes for the hadronic and leptonic boxes in Appendix~\ref{app:FCNCBoxes}. The hadronic and leptonic $\Delta F=1$ EFT contributions can thus be obtained with trivial replacements in the expressions given here.

\subsection{\texorpdfstring{$\Delta F=1$}{DF1} dipole operators}\label{subsec:dipoles}
For dipole-type operators, we define the following effective Lagrangians
\begin{align}
    \mathcal{L}_{\Delta L=1}^{2F}&=-\frac{4G_U}{\sqrt{2}}\frac{1}{16\pi^2}\,W_{21}W_{22}^*\,s_L\, c_Q^2\sum_A \mathcal{C}^e_A\,\mathcal{O}^e_A\, +{\rm h.c.}, \no\\
    \mathcal{L}_{\Delta Q=1}^{2F}&=-\frac{4G_U}{\sqrt{2}}\frac{1}{16\pi^2}\,W_{12}^*W_{22}\,s_Q\, c_L^2 \no\\
  & \qquad  \times \sum_A \left[\mathcal{C}^u_A\,\mathcal{O}^u_A + 
 \mathcal{C}^d_A\,\mathcal{O}^d_A\, \right] +{\rm h.c.}\,,
    \label{eq:Ldipoles}
\end{align}
with the dipole operators defined as
\bea
    \mathcal{O}^u_A &=& \bar q_L^2 \sigma_{\mu\nu} \hat F_A^{\mu\nu}  u_R^3 \tilde H~,  \no  \\
    \mathcal{O}^d_A &=& \bar q_L^2 \sigma_{\mu\nu} \hat F_A^{\mu\nu} d_R^3 H~,  \no  \\
    \mathcal{O}^e_A &=& \bar \ell_L^2 \sigma_{\mu\nu} \hat F_A^{\mu\nu} \ell_R^3 H~.  
\eea
where $\tilde H=i\sigma_2 H^*$, the Higgs vev is normalized such that $\langle H^\dagger H \rangle  =  v^2/2$ with $v\approx 246$~GeV, and
\bea
&& \hat F_B^{\mu\nu} = g_Y\,B^{\mu\nu}~, \qquad \hat F_W^{\mu\nu} = g_2\,T^I W^{I\,\mu\nu}~, \no \\
&&  \hat F_G^{\mu\nu} = g_s\,T^a G^{a\,\mu\nu}~.
\eea
We compute the Wilson coefficients to first order in the (third-generation) SM Yukawa couplings and, consistently with the semileptonic amplitudes discussed above, we neglect flavor-violating effects from the CKM matrix. In this limit, the Wilson coefficients for the $\Delta Q=1$ down-type operators can be written in the following general form
\begin{align}
\mathcal{C}^d_A  &=  -\frac{ y_b}{ 2} \left[ Q^A_{\ell_L}\, G_1 (x_L) + Q^A_{U}\,G_2 (x_L) \right] \no\\
&\quad +  \frac{y_\tau }{ 2 }  \frac{ W_{21} }{   W_{12}^*W_{22} c_L^2 }\, \frac{3}{2} Q^A_{U}\,,
\label{eq:CAd}
\end{align}
where 
\begin{align}
Q^B_{\ell_L}&= -\frac{1}{2}\,, &  
Q^W_{\ell_L}&= 1\,, &
Q^G_{\ell_L}&= 0\,, \no\\
Q^B_U&= \frac{2}{3}\,,&
Q^W_U&= 0\,, &
Q^G_U&= 1\,.
\end{align}
The loop functions 
\begin{align}
G_1(x)&=x\left[\frac{2 - 5 x}{2 (x-1)^4}\,\ln x-\frac{4 - 13 x + 3 x^2}{4 (x-1)^3}\right]\,,\no\\
G_2(x)&=x\left[\frac{4 x-1}{2 (x-1)^4}\,x\ln x+\frac{2 - 5 x - 3 x^2}{4 (x-1)^3}\right]\,,
\end{align}
vanish for $x\to 0$ and approach $G_1 (1) \to -11/24$ and $G_2(1) \to -5/24$ in the $x\to 1$ limit. The separate contributions from each loop diagram are reported in Appendix~\ref{app:Dipoles}. The expression of the (phenomenologically less interesting) coefficient $C^u_A$ is obtained from $C^d_A$ replacing $y_b$ with $y_t$, and $y_\tau$ with the third generation neutrino Yukawa coupling $y_\nu$. In the lepton case we find
\begin{align}
\mathcal{C}^e_A  &=  -\frac{ y_\tau}{ 2} N_c \left[ Q^A_{q_L}\, G_1 (x_Q) + Q^A_{U}\, G_2 (x_Q) \right] \no \\
&\quad+ \frac{y_b }{ 2 } N_c \frac{ W_{12}^* }{   W_{21}W_{22}^* c_Q^2 } 
\, \frac{3}{2} Q^A_{U} , \quad
 \label{eq:WCdipoles}
\end{align}
where $Q^B_{q_L}= \frac{1}{6}$, $Q^W_{q_L}= 1$, and $N_c=3$ is the number of colors of the particles in the loop. Due to the colorless nature of the leptons, there is no gluon-dipole operator.

For completeness, we note that the coefficients of the photon-dipole operators
\be
    \mathcal{O}^d_\gamma = e\, \bar q_L^2 \sigma_{\mu\nu} F^{\mu\nu} d_R^3 H\,,  \quad
    \mathcal{O}^e_\gamma = e\, \bar \ell_L^2 \sigma_{\mu\nu}  F^{\mu\nu} \ell_R^3  H\,,  
\ee
which are particularly interesting from the phenomenological point of view, can be obtained 
by the coefficients above as $\mathcal{C}^{d(e)}_\gamma =\mathcal{C}^{d(e)}_B -\mathcal{C}^{d(e)}_W/2$ or, equivalently, by using 
\eqref{eq:CAd} and 
\eqref{eq:WCdipoles},  with $Q^A_f$ being the electric charges of the corresponding states.

\subsection{\texorpdfstring{$\Delta F=2$}{DF2} hadronic operators}\label{subsec:DF2}
We write the effective Lagrangian for $\Delta Q=2$ transitions as
\begin{align}
    \mathcal{L}_{\Delta Q=2}&=-\frac{4G_U}{\sqrt{2}}\frac{\alpha_4}{4\pi}\,(W_{12}^*W_{22}\,s_Q\, c_L^2)^2\no\\
    &\quad\times F_{\Delta F=2}(x_L)\,(\bar q_L^2\gamma_\mu q_L^3)^2 +{\rm h.c.}\,.
\end{align}
 The loop function 
\begin{align}\label{eq:DF2Box}
F_{\Delta F=2}(x)&=\frac{1}{2}\frac{B_{qq}^{1221}}{(W_{12}^*W_{22}\, c_L^2)^2}\no\\
&=\frac{x (x + 4) (x^2 - 1)}{8 (x - 1)^3}\left[\frac{1}{2}+\frac{x\ln x}{1-x^2}\right]\,,
\end{align}
is such that $F_{\Delta F=2}(1)= 5/48$ and, for small $x$, 
\be
F_{\Delta F=2}(x) = \frac{1}{4} x + \mathcal{O}(x^2)~.
\label{eq:F2smallx}
\ee
The expressions for the individual contributions from each diagram are reported in Appendix~\ref{app:FCNCBoxes}. Note that the loop function in~\eqref{eq:DF2Box} does not agree with the expression in~\cite{DiLuzio:2018zxy,Cornella:2019hct}, which was obtained by rescaling the $W$ box contribution. As we already mentioned in Section~\ref{subsec:FCNCgen}, the different fermion mixing structure of the model compared to the SM does not allow for a naive rescaling of the SM amplitudes. Adopting the same normalization, the loop function  in~\cite{DiLuzio:2018zxy,Cornella:2019hct}  has the same  $x\to 0$ behavior as $F_{\Delta F=2}(x)$ in (\ref{eq:F2smallx}), but a steeper raise for larger $x$ values, reaching 3/16 for $x=1$. As a result, we deduce that the bound on $x_L$ derived in~\cite{DiLuzio:2018zxy,Cornella:2019hct} from $B_s$ mixing is slightly overestimated. 

We finally note that an analogous expression for $\Delta L=2$ transitions is found by replacing $(W_{12}^*W_{22}\,s_Q\, c_L^2)^2\to N_c\,(W_{21}W_{22}^*\,s_L\, c_Q^2)^2$, with the same loop function but with $x_Q$ as argument instead of $x_L$.

\section{Phenomenological implications}\label{sec:pheno}

\subsection{\texorpdfstring{$b\to s\nu\nu$}{b->snunu} transitions}
Before electroweak (EW) symmetry breaking, the part of the effective Lagrangian relevant for $b\to s\nu\nu$ decays is 
\begin{align}\label{eq:bsnn_SMEFT}
    \mathcal{L}_{\rm eff}  \supset -\frac{4G_U}{\sqrt{2}}  \left[\mathcal{C}_{\ell q}^{3333}  \mathcal{O}_{\ell q}^{3333} + \frac{\alpha_4}{4\pi}\, \beta_{23} \,\mathcal{C}_{\ell q}^{3323}  \mathcal{O}_{\ell q}^{3323}   \right]+{\rm h.c.}\,,
\end{align}
where  $\beta_{23}=W_{12}^*W_{22}\,s_Q\, c_L^2$. At the matching scale, $\mathcal{C}_{\ell q}^{3333}$ can be decomposed as\footnote{In the notation of~\cite{Fuentes-Martin:2019ign}, $\mathcal{C}_{\ell q}^{3333}=\mathcal{C}_{\ell q}^{(1)}$.}
\be
\mathcal{C}_{\ell q}^{3333} =  -\frac{1}{4 \xZ } \left[ 1 +  \frac{\alpha_4}{4\pi}\,   \delta \mathcal{C}_{\ell q}^{\rm NLO} (\xZ) \right]\,,
\ee
with $\delta \mathcal{C}_{\ell q}^{\rm NLO} (1) \approx 8$~\cite{Fuentes-Martin:2019ign}. We have checked explicitly using DsixTools (based on the Renormalization Group Evolution (RGE) equations in~\cite{Jenkins:2013zja,Jenkins:2013wua,Alonso:2013hga}) that RGE effects in $\mathcal{C}_{\ell q}^{3333} (\mu=m_Z)$, including the mixing with the flavor-conserving leptoquark mediated operator, are below $20\%$.

\begin{figure*}[t]
    \centering
    \includegraphics[width=0.45\textwidth]{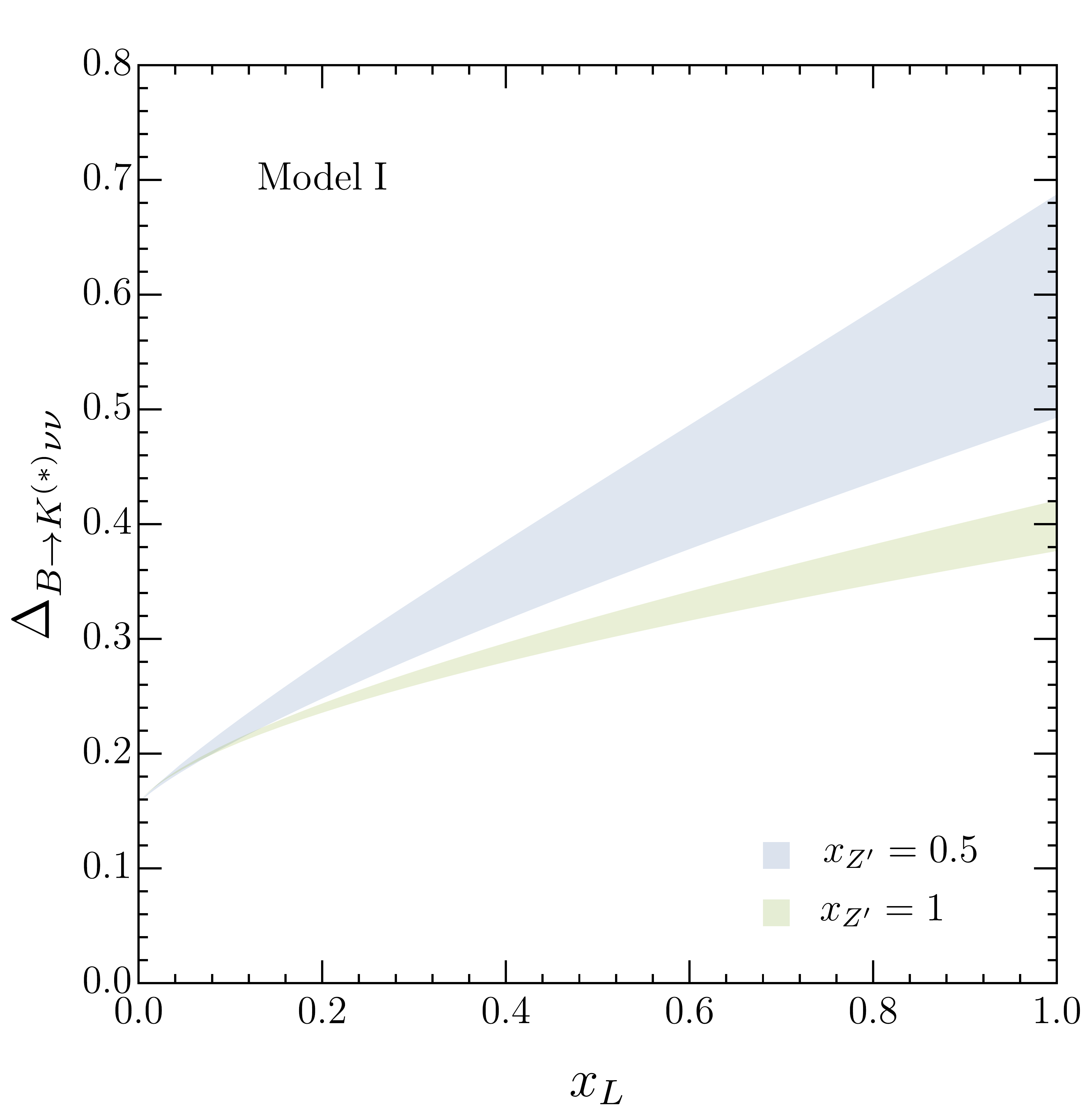}
    \qquad\qquad
    \includegraphics[width=0.45\textwidth]{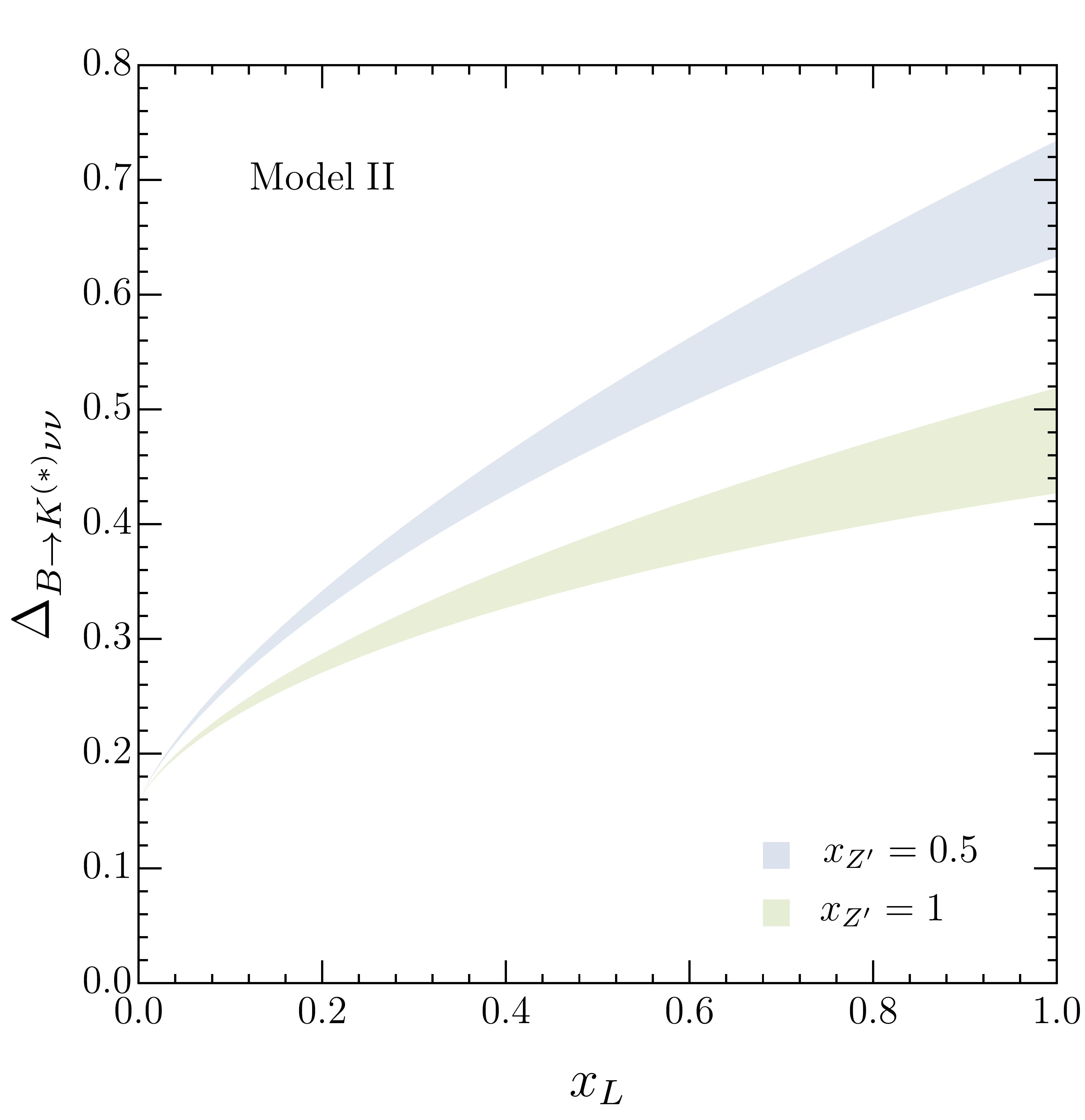}
    \caption{Model predictions for the modifications of $\cB(B\to K^{(*)} \nu\nu)$ relative to the SM predictions, see Eq.~(\ref{eq:DeltaKnn}). The bands are due to the variation of radial masses in the range $m_R\in [1,2\pi]\times m_U$ (with $f_R=0$ in model II). See main text for more details.  }
    \label{fig:B2Knunu}
\end{figure*}

After EW symmetry breaking, we can project the contributions of $ \mathcal{O}_{\ell q}^{3333}$ and $ \mathcal{O}_{\ell q}^{3323}$ onto the coefficients of the the operators $(\bar s_L \gamma_\mu b_L) (\bar\nu_{\ell}\gamma^\mu \nu_{\ell})$, that we normalize as in the SM,
\begin{equation}
\cL_{\rm eff} = -\frac{4 G_F}{\sqrt{2}}\frac{\alpha_w}{2\pi}\, V_{tb}V_{ts}^*\,  C_{\ell}\, 
(\bar s_L \gamma_\mu b_L) (\bar\nu_{\ell}\gamma^\mu \nu_{\ell})\,,
\label{eq:bsnn_LEET}
\end{equation}
where $\alpha_w =  \alpha/ s^2_w = g_2^2/(4\pi)$  and $V_{ij}$ are CKM matrix elements. 
Neglecting the tiny contributions suppressed by light quark masses, the 
lepton-universal SM contribution read
\begin{equation}
C^{{\rm SM}}_{\ell} = X_t\,,
\label{eq:bsnn_CSM}
\end{equation}
where $X_t = 1.48\pm 0.01$~\cite{Buchalla:1998ba}. Taking into account the contribution 
of (\ref{eq:bsnn_SMEFT}) after diagonalizing the Yukawa couplings, we get 
$C_{\ell} \approx  C^{{\rm SM}}_{\ell} $ for $\ell = e,\mu$ and 
\be
C_{\tau}  \approx  C^{{\rm SM}}_{\tau}  \left[  1  +  \rho \,    \Delta C_{\tau} \right]\,,
\ee
where  
\begin{align}\label{eq:DeltaCtau}
\Delta C_{\tau} &=\frac{s_b}{ V_{tb} V_{ts}^*}\,\mathcal{C}_{\ell q}^{3333} +  \frac{ \beta_{b\tau}^*\beta_{s \tau} }{ V_{tb} V_{ts}^*}\left(\frac{\alpha_4}{4\pi}\,\mathcal{C}_{\ell q}^{3323}+\mathcal{C}_U^{\rm RGE}\right)\,,\no\\
\rho &= \frac{2\pi}{\alpha_w X_t} \frac{G_U}{G_F}  = 1.26 \times \left(\frac{C_U}{0.01}\right)\,.
\end{align}
Here, $s_b$ is the 2-3 mixing from the left-handed diagonalization of $Y_d$, defined as in~\cite{Cornella:2019hct,Fuentes-Martin:2019mun}, and we have neglected terms of $\cO(s_b^2)$. Employing the notation of~\cite{Cornella:2019hct}, we further identified $C_U= G_U/G_F$ and $\beta_{23}=\beta_{b\tau}^*\beta_{s \tau}+\mathcal{O}(s_{L,d}^2)$, and neglected the terms of $\mathcal{O}(s_{L,d}^2)$. Finally, $\mathcal{C}_U^{\rm RGE}$ encodes the RGE-induced contribution from the flavor-violating tree-level leptoquark mediated operator. Using DsixTools~\cite{Celis:2017hod} and setting $m_U=4~\mathrm{TeV}$, we find
\begin{align}
\mathcal{C}_U^{\rm RGE}\approx-0.058\,.
\end{align}
With this notation, we can write 
\be
\frac{\cB(B\to K^{(*)} \nu\nu )}{\cB(B\to K^{(*)} \nu\nu )_{\rm SM}}=  1+\frac{2}{3}\, \rho\,  \Delta C_{\tau} + \frac{1}{3}\,\rho^2\,\Delta C_{\tau}^2\,.
\ee
Of the two contributions to $ \Delta C_{\tau}$ in \eqref{eq:DeltaCtau}, the one proportional to $s_b$ can induce at most a $\pm 3\%$ correction to $\cB(B\to K^{(*)} \nu\nu )$: the value of $s_b$ is indeed severely constrained by the tree-level $\Zp$ and $\Gp$ contributions to $B_s$ mixing, which imply $|s_b| \lsim 0.1 \times |V_{ts}|$~\cite{Bordone:2017bld,DiLuzio:2017vat,DiLuzio:2018zxy,Cornella:2019hct}. The contribution proportional to $\beta_{23}\approx\beta_{s\tau}\beta_{b\tau}^*$ can be larger, yielding up to $\mathcal{O}(60\%)$ corrections to the $\cB(B\to K^{(*)} \nu\nu )$ SM value. Moreover, the sign of the correction is unambiguously connected to the sign of the new physics contributions to $R_{D^{(*)}}$. More precisely, an enhancement of the $R_{D^{(*)}}$ ratios requires a  positive $\beta_{23}$ that, in turn, implies an enhancement also in $\cB(B\to K^{(*)} \nu\nu)$. 

In Fig.~\ref{fig:B2Knunu} we plot 
\be\label{eq:DeltaKnn}
\Delta_{B\to K^{(*)}\nu\nu}\equiv \frac{ \cB(B\to K^{(*)} \nu\nu ) }{    
\cB(B\to K^{(*)} \nu\nu )_{\rm SM}} -1\,,
\ee 
as a function of $x_L$ setting $\beta_{23} = 3|V_{ts}|$, $C_U = 0.01$ and $g_4=3$, which are natural benchmark values to fit $R_{D^{(*)}}$ while avoiding direct searches~\cite{Fuentes-Martin:2019mun}. Changing $\beta_{23}$ and $C_U$ leads to uniform linear re-scaling of the plot
\be
\Delta_{B\to K^{(*)}\nu\nu} \to  \Delta_{B\to K^{(*)}\nu\nu}^{\rm [Fig.~1]} 
\times \left( \frac{C_U}{0.01} \right) \left( \frac{\beta_{23}}{ 3|V_{ts}| }  \right)~.
\ee 
The value of $\Delta_{B\to K^{(*)}\nu\nu}$ at $x_L=0$ corresponds to the contribution of $\mathcal{C}_U^{\rm RGE}$ in (\ref{eq:DeltaCtau}), whereas the growth with $x_L$ is due to $\mathcal{C}_{\ell q}^{3323}$. As a result, a change of $g_4$ would rescale only the latter contribution. It is worth noting that the Belle II Collaboration should be able to measure $\cB(B\to K^{(*)} \nu\nu )$ with a 10\% error, assuming the SM value~\cite{Kou:2018nap}, and thus should be able to probe  most of the  parameter space of the model relevant to fit the $B$-physics anomalies. 

\subsection{\texorpdfstring{$b\to c\tau\nu$}{b->ctaunu} transitions}

In this section we evaluate the modifications of $b\to c\tau\nu$ decay amplitudes, and their impact in $R_D$ and $R_{D^*}$, with respect to the NLO effects estimated in~\cite{Fuentes-Martin:2019ign,Fuentes-Martin:2020luw} in the limit of minimal field content. In the rest of the section we refer to these previous works as Ref.~I~\cite{Fuentes-Martin:2019ign} and II~\cite{Fuentes-Martin:2020luw}.

Before EW symmetry breaking, the effective Lagrangian relevant to charged-current transitions can be decomposed as  
\begin{align}
\mathcal{L}_{\rm CC}=-\frac{4G_U}{\sqrt{2}}\left[\mathcal{C}_{LL}^U\,\mathcal{O}_{LL}^U+(\mathcal{C}_{LR}^U\,\mathcal{O}_{LR}^U+\mathrm{h.c.})\right]\,,
\label{eq:L_CC0}
\end{align}
where we have left the flavor indices implicit, and the operators are defined as 
\begin{align}
[\mathcal{O}_{LL}^U]^{\alpha i j  \beta} &=(\bar \ell^\alpha_L\gamma_\mu q^i_L)(\bar q^j_L\gamma^\mu\ell^\beta_L)\,, \nonumber\\
[\mathcal{O}_{LR}^U]^{\alpha i j  \beta} &=(\bar \ell^\alpha_L\gamma_\mu q^i_L)(\bar d^j_R\gamma^\mu e^\beta_R)\,.
\end{align}
Restricting the attention to $b\to c\tau\nu$ decays, quark flavor indices assume the values 3 and 2, whereas the lepton flavor indices are always third generation (in close analogy to the $b\to s\nu\nu$ case discussed above).

At the matching scale, the relevant Wilson coefficients can be decomposed as
\begin{align}
[\mathcal{C}_{LL}^U]^{3333} &=|W_{11}|^2 \left[1+\frac{\alpha_4}{4\pi}\, \delta\mathcal{C}_{LL}^{(4),33} + 
\frac{\alpha_s}{4\pi}\,\delta\mathcal{C}_{LL}^{(s),33} \right]\,,\nonumber\\
[\mathcal{C}_{LR}^U]^{3333}&=W_{11}^*\, e^{i\phi_{LR}}
\left[1+\frac{\alpha_4}{4\pi}\, \delta\mathcal{C}_{LR}^{(4),33} + 
\frac{\alpha_s}{4\pi}\,\delta\mathcal{C}_{LR}^{(s),33} \right]\,, \nonumber\\
[\mathcal{C}_{LL}^U]^{3233} &= -W_{21}^*  W_{11} s_Q \!\left[1+\frac{\alpha_4}{4\pi}\, \delta\mathcal{C}_{LL}^{(4),32} + 
\frac{\alpha_s}{4\pi}\,\delta\mathcal{C}_{LL}^{(s),32} \right]\,,\nonumber\\
[\mathcal{C}_{LR}^U]^{3233}&= -W_{21}^* \, s_Q\, e^{i\phi_{LR}}
\left[1+\frac{\alpha_4}{4\pi}\, \delta\mathcal{C}_{LR}^{(4),32} + 
\frac{\alpha_s}{4\pi}\,\delta\mathcal{C}_{LR}^{(s),32} \right]\,,
\label{eq:WC_CC1}
\end{align}
where $\phi_{LR}$ parametrizes the arbitrary relative phase between left- and right-handed currents, related to the embedding of SM quark and leptons in $SU(4)$ multiplets~\cite{Bordone:2017bld}. The first term in all the expressions above corresponds to the tree-level contribution that, compared to Ref.~I and II, is modulated by a combination of $W$ entries also in the flavor-conserving case.  
 
The NLO corrections can be further decomposed into a factorizable contribution due to the renormalization of $g_4$ (under both $\alpha_4$ and $\alpha_s$ corrections) and non-factorizable finite contributions due to box amplitudes and non-universal vertex corrections. In order to follow the approach adopted in  Ref.~I and II as closely as possible, we renormalize $g_4$ from the on-shell inclusive decay width of the LQ into a $\tau$ lepton and any quark species, that we denote as $\Gamma^U_\tau$. In the absence of high-energy observables sensitive to $W_{ij}$, we treat $W_{21}$ (and correspondingly $W_{11}$) as an effective low-energy parameter that we do not need to normalize.

By construction, the $\alpha_s$ corrections are flavor blind and can be directly extracted from the result in Ref.~II. Summing factorizable and non-factorizable contributions, and assuming the custodial limit for the vector masses, yields
\be
\delta\mathcal{C}_{LL}^{(s),ij} \approx 2.65~, \qquad 
\delta\mathcal{C}_{LR}^{(s),ij} (\mu=m_{U}) \approx 7.15~.
\ee

As far as the leading $\alpha_4$ corrections are concerned, the renormalization of $g_4$ proceeds as in Ref.~I and II. The unitarity of the $W$ matrix ensures that the finite vertex corrections are independent of $W$ to a good approximation. The residual $W$ dependence of $\Gamma^U_\tau$, proportional to $|W_{21}|^2$, vanishes in the limit $m^2_{Q,L} \ll m^2_V$ and is expected to be subleading. This subleading contribution is model dependent and we neglect it in the following. We can thus decompose the $\alpha_4$ NLO corrections as
\begin{align} 
\delta\mathcal{C}_{LL(R)}^{(4),ij} =\,  &
\delta\mathcal{C}^{(4)}|^{\rm I}_{\rm F} +
\delta\mathcal{C}_{LL(R)}^{(4)}|^{\rm I}_{\rm NF} \no\\
\quad   +\,  & \delta\mathcal{C}^{(4)}|^{\rm VL}_{\rm F} +
\delta\mathcal{C}_{LL(R)}^{(4),ij}|^{\rm VL}_{\rm NF} \,.
\end{align}
Here the subscript I refers to the flavor-blind 
result obtained in Ref. I that, in the custodial limit 
for the vector masses, yields
\be
\delta\mathcal{C}^{(4)}|^{\rm I}_{\rm F} \approx 0~, \qquad 
 \delta\mathcal{C}_{LL(R)}^{(4)}|^{\rm I}_{\rm NF} \approx 
\frac{11}{4} \left( \frac{29}{4} \right)~.
\ee
where the subscript F (NF) denotes the factorizable (non-factorizable) contributions. The factorizable contribution due to vector-like quarks, $\delta\mathcal{C}^{(4)}|^{\rm VL}_{\rm F}$, corresponds to the two-point function corrections that can be found in Appendix~\ref{app:Uself}. This is the only effect due to these additional degrees of freedom that does not vanish in the $x_{Q,L} \to 0$ limit. We find that this contribution yields an $\mathcal{O}(5\%-10\%)$ reduction of the WCs, for fixed on-shell coupling $g_4=3$. As far as non-factorizable corrections are concerned, $\delta\mathcal{C}_{LL(R)}^{(4),ij}|^{\rm VL}_{\rm NF}$, we neglect the contributions generated from the vertex, consistently with what we did with the $g_4$ renormalization, and consider only the box contributions. Given the results in Ref. I, the vertex contributions are expected to be numerically subleading compared to the box amplitudes. The complete expressions for the box amplitudes can be found in Appendix~\ref{app:CCBoxes}. In the custodial limit for the vector masses, we have
\begin{align}
\delta\mathcal{C}_{LL(R)}^{(4),33}|^{\rm VL}_{\rm NF}&\approx0\,,\no\\
\delta\mathcal{C}_{LL}^{(4),32}|^{\rm VL}_{\rm NF}  &\approx\frac{15}{8}\,c_Q^2\,\frac{x_Q}{(1-x_Q)^2}\left(1-x_Q+\ln x_Q\right)\,,\no\\
\delta\mathcal{C}_{LR}^{(4),32}|^{\rm VL}_{\rm NF}&\approx \frac{45}{8}\,c_Q^2\,\frac{x_Q}{(1-x_Q)^2}\left(1-x_Q+\ln x_Q\right)\,,
\end{align}
which in the $x_Q,c_Q\to 1$ limit yield 
\be
\delta\mathcal{C}_{LL}^{(4),32}|^{\rm VL}_{\rm NF}=-15/16\,, \quad \delta\mathcal{C}_{LR}^{(4),32}|^{\rm VL}_{\rm NF}=-45/16\,.
\ee

The ratio between LL and LR effective operators is of phenomenological relevance, since it affects the relative weight of scalar and vector contributions to $R_{D}$ and  $R_{D^*}$. At the tree level, this ratio is completely determined by $W_{11}$ and the phase $\phi_{LR}$. At NLO accuracy it gets modified by the non-factorizable corrections and becomes flavor dependent. To parametrize this effect, we define the WC ratios at the matching scale 
\begin{align}
\rho_{LR}^{33}\equiv\frac{[\mathcal{C}_{LR}^U]^{33}}{[\mathcal{C}_{LL}^U]^{33}}&=\frac{e^{i\phi_{LR}}}{W_{11}}\,\big[1+\frac{\alpha_4}{4\pi}(\delta\mathcal{C}_{LR}^{(4)}|^{\rm I}_{\rm NF}-\delta\mathcal{C}_{LL}^{(4)}|^{\rm I}_{\rm NF})\big]\no\\
&\quad+\frac{\alpha_s}{4\pi}(\delta\mathcal{C}_{LR}^{(s),33}-\delta\mathcal{C}_{LL}^{(s),33})\,,\no\\
\rho_{LR}^{32}\equiv \frac{[\mathcal{C}_{LR}^U]^{32}}{[\mathcal{C}_{LL}^U]^{32}}&=\frac{e^{i\phi_{LR}}}{W_{11}}\,\big[\rho_{LR}^{33}+\frac{\alpha_4}{4\pi}(\delta\mathcal{C}_{LR}^{(4),32}|^{\rm VL}_{\rm NF}\no\\
&\quad-\delta\mathcal{C}_{LL}^{(4),32}|^{\rm VL}_{\rm NF})\big]\,.
\end{align}
At fixed $g_4=3$, we have $\rho_{LR}^{33}\approx 1.29\,e^{i\phi_{LR}}/W_{11}$ and $\rho_{LR}^{32}/\rho_{LR}^{33}\in[1.0,0.92]$ for $c_Q=1$ and $x_Q\in[0,1]$. 

We have now collected all the ingredients to provide a description of the LQ contributions to the $R_{D^{(*)}}$ ratios at NLO accuracy. Expressing the quark fields in terms of mass eigenstates (after EW symmetry breaking), and evolving the effective operators down to $\mu=m_b$, we obtain
\begin{align}
\Delta R_X&\equiv\frac{R_X}{R_X^{\rm SM}}-1\no\\
&\approx2C_U\,\Big[[\mathcal{C}_{LL}^U]^{3333}\,(1-\eta_S\,c_X\,\rho_{LR}^{33})\no\\
&\quad+[\mathcal{C}_{LL}^U]^{3233}\,\frac{V_{cs}}{V_{cb}}\,(1-\eta_S\,c_X\,\rho_{LR}^{32})\Big]\,,
\end{align}
where $X=D,{D^*}$. Here $\eta_S$ is the factor encoding the RGE evolution of $\mathcal{O}_{LR}^U$, which for $m_U=4~\mathrm{TeV}$ assumes the value $\eta_S\approx 1.8$~\cite{Celis:2017hod}. The coefficients $c_X$ encode the ratios of the hadronic matrix elements of scalar and vector operators in the two modes. According to~\cite{Feruglio:2018fxo,Fajfer:2012vx}, they are given by $c_D\approx1.5$ and $c_{D^*}\approx0.14$.

\section{Conclusions}\label{sec:conclusions}

In this paper we have presented a systematic analysis of the impact of vector-like fermions, beyond the tree level, in models based on the (flavor non-universal) $SU(4)\times SU(3)^\prime\times SU(2)_L\times U(1)_X$ gauge group. The inclusion of such heavy fields in this class of models is necessary for a successful phenomenological description of the SM spectrum at low-energies, in particular to describe masses and mixing angles for the light generations~\cite{Bordone:2017bld,Cornella:2019hct,Greljo:2018tuh}. Vector-like fermions are also a key ingredient to enhance the $3-2$ flavor mixing in the effective coupling of the TeV-scale LQ field to SM fermions, providing a better fit to the charged-current  $B$ anomalies~\cite{DiLuzio:2018zxy}. We have considered two possible embeddings of the vector-like fermions into the model, both satisfying these phenomenological requirements. Interestingly, most of the conclusions we have derived are, to a large extent, independent of the specific embedding. 

The new sources of flavor symmetry breaking due to the additional mass terms associated to the vector-like fermions lead to non-vanishing FCNC amplitudes that are not present in the minimal version of the model. We have elucidated the origin of this phenomenon in general terms, and we have systematically analyzed the matching conditions for FCNC semileptonic, dipole and $\Delta F=2$ operators. Using these results, combined with previous NLO results in~\cite{Fuentes-Martin:2019ign}, we present the first complete analysis of the impact of the $U_1$ leptoquark in $B\to K^{(*)} \nu\nu$ decays beyond the tree level. As shown in Fig.~\ref{fig:B2Knunu}, the branching ratios of these rare modes are unambiguously predicted to be enhanced by $10\%$ to $50\%$ in the parameter region of the model providing a good fit to the $B$-physics anomalies.
 
The inclusion of vector-like fermions leads also to sizable NLO effects in amplitudes which are non-vanishing already at the tree-level, such as charged-current semileptonic transitions. Extending our previous works~\cite{Fuentes-Martin:2019ign,Fuentes-Martin:2020luw}, we have analyzed these additional NLO effects. Using these results, we have derived phenomenological expressions of the $R_{D^{(*)}}$ ratios, in terms of the model parameters, which include all the relevant corrections at $\cO(\alpha_4)$ and $\cO(\alpha_s)$. These results will allow us to perform precise compatibility tests of the $B$-physics anomalies, if confirmed as clear signals of physics beyond the SM, with the predictions of 4321 models.

\section*{Acknowledgments}

This project has received funding from the European Research Council (ERC) under the European Union's Horizon 2020 research and innovation programme under grant agreement 833280 (FLAY), and by the Swiss National Science Foundation (SNF) under contract 200021-175940. The work of J.F. was also supported in part by the Generalitat Valenciana under contract SEJI/2018/033.

\appendix

\section{Vector-like fermion implementations}\label{app:VLmodels}

As discussed in~\ref{sec:VLferm}, there are several possible implementations for the massive fermions. In this appendix, we discuss in more detail the two realizations corresponding to model I and II in Table~\ref{tab:VLcontent}. We also complete the discussion in~\ref{sec:VLferm} by including Goldstone boson and (or) physical scalar interactions with fermions for each implementation.

\subsection{Model I}

This model consists of a simplified version of the composite model in~\cite{Fuentes-Martin:2020bnh}, where a single vector-like family is included. In this implementation, the vector-like mass and fermion mixing terms are given by ($i=2,3$)
\begin{align}\label{eq:extLag}
\mathcal{L}_{\rm mix}&=\lambda^i_Q\,\bar\Psi_L^{\prime i}\,\Omega_3^\dagger\,Q_R+\lambda^i_L\,\bar\Psi_L^{\prime i}\,\Omega_1^\dagger\,L_R\no\\
&\quad+M_q\,\bar q_L^{\prime\,2}\,Q_R+M_\ell\,\bar\ell_L^{\prime\,2}\,L_R+{\rm h.c.}\,,
\end{align}
where $\Psi_L^{\prime\,i}=(\Psi_L^{q\,\prime\,i}\;\Psi_L^{\ell\,\prime\,i})$ with $\Psi_L^{q,\ell\,\prime}$ defined as in~\eqref{eq:Psiprime}, and with $\Omega_{1,3}$ as in~\eqref{eq:Omega13}. In the composite model in~\cite{Fuentes-Martin:2020bnh}, one has $\omega_1=\omega_3$, so the vevs of $\Omega_{1,3}$ preserve the custodial $SU(4)_V$ symmetry. Moreover, only the Goldstone and vev part of $\Omega_{1,3}$ is the same as in~\eqref{eq:Omega13}, while the physical scalars, together with other composite resonances, are expected to have masses around the compositeness scale, $\Lambda\approx 4\pi\omega_{1,3}$, much larger than the heavy gauge boson masses. However, to illustrate the effect of the radial modes in the computation of the FCNCs, we leave this model general by treating $\omega_1$ and $\omega_3$ as independent parameters and we keep the leptoquark radial in $\Omega_{1,3}$.

After $\Omega_{1,3}$ acquires a vev, it is straightforward to write the fermion mass terms in the form of~\eqref{eq:MassMix} (see also~\eqref{eq:VLMass14}), with 
\begin{align}\label{eq:MassI}
    M_q&=
    \begin{pmatrix}
    M_q\\[2pt]
    \lambda_Q \frac{\omega_3}{\sqrt{2}}
    \end{pmatrix}
    \,,&
    M_\ell&=
    \begin{pmatrix}
    M_\ell\\[2pt]
    \lambda_L \frac{\omega_1}{\sqrt{2}}
    \end{pmatrix}
    \,.
\end{align}
Moving to the $SU(4)$ basis defined in~\eqref{eq:SU4basis}, the Lagrangian in~\eqref{eq:extLag} can be rewritten as
\begin{align}\label{eq:Lmix}
\mathcal{L}_{\rm mix}&= m_Q\,\bar Q_LQ_R+m_L\,\bar L_LL_R+\mathcal{L}_{\rm GB}+\mathcal{L}_{\rm rad}\,,
\end{align}
with
\begin{align}\label{eq:GBfermInt}
\mathcal{L}_{\rm GB}&=\frac{g_4}{\sqrt{2}}\,\phi_U\!\left(\frac{m_L}{m_U}\,c_L\,W_{i2}\, \mathcal{\bar Q}_L^i\, L_R-\frac{m_Q}{m_U}\,c_Q\,W_{2i}\,\bar Q_R\, \mathcal{L}_L^i\right)\no\\
&\quad+i\frac{g_4}{2\sqrt{6}}\,\phi_{Z^\prime}\left(\frac{m_Q}{m_{Z^\prime}}\,c_Q\,\mathcal{\bar Q}_L^2 Q_R-3\frac{m_L}{m_{Z^\prime}}c_L\, \mathcal{\bar L}_L^2 L_R\right)\no\\
&\quad+ig_4\,\phi_{G^\prime}^a\frac{m_Q}{m_{G^\prime}}\,c_Q\,\mathcal{\bar Q}_L^2\, T^a Q_R+{\rm h.c.},\\
\mathcal{L}_{\rm rad}&\supset\frac{g_4}{\sqrt{2}}\,h_U\!\left(\frac{m_L}{m_U}\,c_L\,W_{i2}\,\cot\beta\, \mathcal{\bar Q}_L^i\, L_R\right.\no\\
&\quad\left.+\frac{m_Q}{m_U}\,c_Q\,W_{2i}\,\tan\beta\,\bar Q_R\, \mathcal{L}_L^i\right)\,.
\end{align}
and where, in the radial interactions, we included only the leptoquark interactions. 

\subsection{Model II}

This model consists of a simplified version of the one in~\cite{Cornella:2019hct}, with only one vector-like family. Since in this implementation $\chi_R$ is an $SU(4)$ multiplet, a new source of $SU(4)$ breaking beyond the $\Omega_{1,3}$ vevs is needed to generate a mixing between $SU(4)$ flavor states. This can be obtained from the vev of a new scalar field, $\Omega_{15}$, transforming in the adjoint of $SU(4)$ and singlet under the rest of the 4321 group. Once this new field is introduced, the vector-like mass and fermion mixing terms for this model read ($i=2,3$)
\begin{align}\label{eq:extLag2}
\mathcal{L}_{\rm mix}&=\lambda_{15}^i\,\bar\Psi_L^i\,\Omega_{15}\,\chi_R+M_\chi^i\,\bar\Psi_L^i\,\chi_R\no\\
&\quad+\lambda_q\,\bar q_L^{\prime\,2}\,\Omega_3\,\chi_R+\lambda_\ell\,\bar\ell_L^{\prime\,2}\,\Omega_1\,\chi_R+{\rm h.c.}
\end{align}
where $\Psi_L^{\prime\,i}=(\Psi_L^{q\,\prime\,i}\;\Psi_L^{\ell\,\prime\,i})$ with $\Psi_L^{q,\ell\,\prime}$ defined as in~\eqref{eq:Psiprime}. Since the scalar sector of this model is more complicated, we do not discuss the radial modes here. The Goldstone and vev part of $\Omega_{1,3}$ retain the same form as in~\eqref{eq:Omega13} (with $m_U$ as in~\eqref{eq:MUO15}), while the Goldstone and vev part of $\Omega_{15}$ decompose under the SM group as
\begin{align}
\Omega_{15}=\frac{\omega_{15}}{2\sqrt{6}}
\begin{pmatrix}
\mathbb{1}_{3\times 3} & -4\,\frac{g_4}{\sqrt{2}}\frac{\phi_U}{m_U}\\
-4\,\frac{g_4}{\sqrt{2}}\frac{\phi_U^\dagger}{m_U}& -3\\
\end{pmatrix}
+\dots\,,
\end{align}
where the dots represent radial excitations that we do not consider. The presence of a vev for $\Omega_{15}$ introduces an explicit breaking of the custodial $SU(4)$ symmetry in the gauge boson masses. Indeed, this vev does not affect the $Z^\prime$ and $G^\prime$ masses, but it does change the $U_1$ mass compared to the one given in~\eqref{eq:SU4vectorMasses}. More precisely, we now have
\begin{align}\label{eq:MUO15}
m_U=\frac{g_4}{2}\sqrt{\omega_1^2+\omega_3^2+\frac{4}{3}\,\omega_{15}^2}\,.
\end{align} 

Once more, it is possible to write the fermion mass terms in the form of~\eqref{eq:MassMix} after $\Omega_{1,3,15}$ acquire a vev. Namely,
\begin{align}\label{eq:MassII} 
    M_q&=
    \begin{pmatrix}
    \lambda_q\,\frac{\omega_3}{\sqrt{2}}\\[2pt]
    M_\chi+M_{15}
    \end{pmatrix}
    \,,&
    M_\ell&=
    \begin{pmatrix}
    \lambda_\ell\,\frac{\omega_1}{\sqrt{2}}\\[2pt]
    M_\chi-3\,M_{15}
    \end{pmatrix}
    \,,
\end{align}
where $M_{15}\equiv\lambda_{15}\,\omega_{15}/(2\sqrt{6})$. Note that, in the limit $\omega_{15}=0$, the mass vectors are aligned and the $W=W_q^\dagger\, W_\ell$ matrix becomes the identity. Using the same decomposition as in~\eqref{eq:Lmix}, we now find for the Goldstone boson interactions
\begin{align}
\mathcal{L}_{\rm GB}&=\frac{g_4}{\sqrt{2}}\,\phi_U\left[\left(\frac{m_L}{m_U}\,c_L\,W_{i2}\,\mathcal{\bar Q}_L^i\, L_R-\frac{m_Q}{m_U}\,c_Q\,W_{2i}\,\bar Q_R\, \mathcal{L}_L^i\right)\right.\no\\
&\left.\hspace{1.8cm}+\left(\frac{m_L}{m_U}\,\bar Q_R\, L_L-\frac{m_Q}{m_U}\,\bar Q_L\, L_R\right)\right]\no\\
&\quad-i\frac{g_4}{2\sqrt{6}}\,\phi_{Z^\prime}\left(\frac{m_Q}{m_{Z^\prime}}\,s_Q\,\bar q_L^\prime Q_R-3\frac{m_L}{m_{Z^\prime}}s_L\,\bar \ell_L^\prime L_R\right)\no\\
&\quad-ig_4\,\phi_{G^\prime}^a\frac{m_Q}{m_{G^\prime}}\,s_Q\,\bar q_L^\prime\, T^a Q_R+{\rm h.c.},
\end{align}
with $q_L^\prime=c_Q\,q_L^2+s_Q Q_L$, and analogously for $\ell_L^\prime$ (see~\eqref{eq:Omatrices} for the definition of the mixing angles). Note that the first term in the $\phi_U$ interactions coincides with the one in the previous model, c.f.~\eqref{eq:GBfermInt}. The second term is new and is related to the fact that $\chi_R$ is now charged under $SU(4)$. Also note that, contrary to the previous case, there are no Goldstone couplings to $\phi_{Z^\prime,G^\prime}$ in the limit $s_{q,\ell}\to0$, making manifest the custodial symmetry breaking. The interactions involving the SM fields are however the same in both models.

\subsection{\texorpdfstring{$SU(4)_V$}{SU4V} structure of \texorpdfstring{$W$}{W} and \texorpdfstring{$O_{q,\ell}$}{Oql}}

A complementary (model-independent) view about the mixing matrices $W$ and $O_{q,\ell}$
is obtained by looking at their transformation properties under the $SU(4)_V$ custodial symmetry.
To do so, we rewrite \eqref{eq:MassMix} using a $SU(4)$-invariant notation,
\be
\cL_{\rm mass} = {\bar \xi}_L   \cM^{4}  \chi_R +  \psi^{\prime 2}_L \cM^{1}  \chi_R~,
\ee
with $\xi_L$ defined in \eqref{eq:xidef}.
Since the  two $\cM^{a}$ mix different $SU(4)$ representations with the same right-handed field, 
one of  them necessarily break the $SU(4)$ gauge symmetry ($\cM^{4}$ in model I, and  $\cM^{1}$ in model II). We can further decompose the 
$\cM^{a}$ in the $SU(4)_V$ space as 
\be
\cM^{a} = \frac{1}{4}\left[ \cMn^{a} + \Delta \cM^{a}  T_{Q-L} \right]~, 
\ee
where  $T_{Q-L}  = \frac{3}{2} (T_{B-L} +\frac{1}{3} )$, such that the 
$M^a_{q,\ell}$ defined in 
\eqref{eq:VLMass14} are
\be
M^a_{q,\ell} =  \cMn^{a} \pm \Delta \cM^{a}~.
\ee
From this decomposition we see that, in addition to the breaking of the $SU(4)$ gauge symmetry,
the $\cM^{a}$ can break the $SU(4)_V$ symmetry if  $\Delta \cM^{a} \not=0$.
Finally, since $\cM^{4}$ is a vector in flavor space,
it can give rise to flavor mixing if its $SU(4)_V$ conserving and violating components are not aligned in the   $U(2)_\xi$ flavor space.

The rotation matrices $\tilde W_{q,\ell}$ and $O_{q,\ell}$ are determined by the diagonalization of 
the  $3\times3$ hermitian matrix $\cM_L \cM_L^\dagger$, where $ \cM_L$ is the vector
\be 
\qquad    \cM^T_L  = \left( \cM^{1}, \cM^4_1,  \cM^4_2 \right)~,
\ee
for quark and leptons. 
The matrix  $\cM_L \cM_L^\dagger$ has rank one and is dominated by $\cMn^{4}$. 
To understand how the mixing matrices are related to the breaking of the various symmetries, 
let us consider the basis where $\cMn^{4}$ is aligned to the second generation of $\xi_L$,
\be
\cMn^{4} = \left( \ba{c} 0 \\ M_\chi \ea \right)~,
\ee
and let us consider the limiting case where all the other contributions to  $\cM_L$ 
are small relative to $M_\chi$. Then from the perturbative diagonalization of $\cM_L \cM_L^\dagger$ we obtain
\bea
&& (O_{q,\ell} )_{13}  \approx     \frac{  \cM^{1} \pm \Delta \cM^{1}  } {M_\chi} =  \frac{ M^1_{q,\ell}   } {M_\chi},  \no \\
&& W_{12}  \approx   (\tilde W_{\ell} - \tilde W_{q})_{23} \approx -  2 \frac{ \Delta \cM^4_1 } {M_\chi} = 
 \frac{  (M^4_{\ell})_1 -(M^4_{q})_1   } {M_\chi}.  \no \\
 \label{eq:Deltapert}
\eea
Form this we deduce that 
\begin{itemize}
\item $W\not=1$ can be achieved only with a double breaking of 
 $SU(4)_V$ and the $U(2)_\xi$ flavor symmetry in $\cM^{4}$.
 \item $O_{q,\ell} \not=1$ necessarily require $SU(4)$ breaking,
 involving both $\cM^{1}$ and $\cM^{4}$, but does not require 
 $SU(4)_V$ breaking. If the custodial symmetry is unbroken $O_{q}=O_{\ell}$.
\end{itemize}

\section{Details on the FCNC computations}\label{app:loopDetails}

\subsection{\texorpdfstring{$Z^\prime$}{Zp} and \texorpdfstring{$G^\prime$}{Gp} flavor-changing vertices}

\subsubsection{Contribution from gauge and Goldstone fields}

\begin{figure}[t]
\centering
\includegraphics[width=0.4\textwidth]{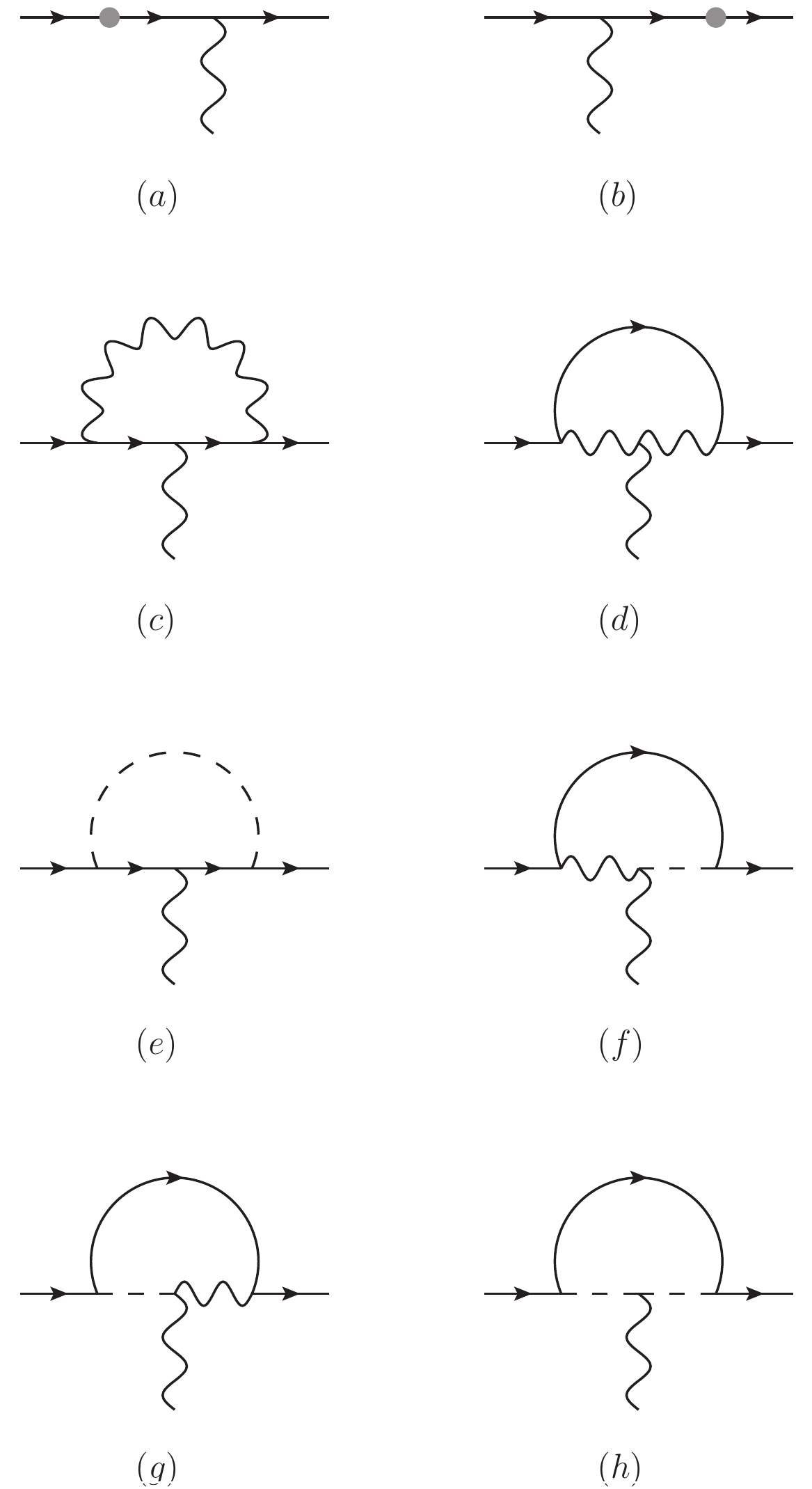}
\caption{One loop diagrams contributing to the $Z^\prime$ and $G^\prime$ flavor-changing vertices.}
\label{fig:vertices}
\bigskip
\includegraphics[width=0.4\textwidth]{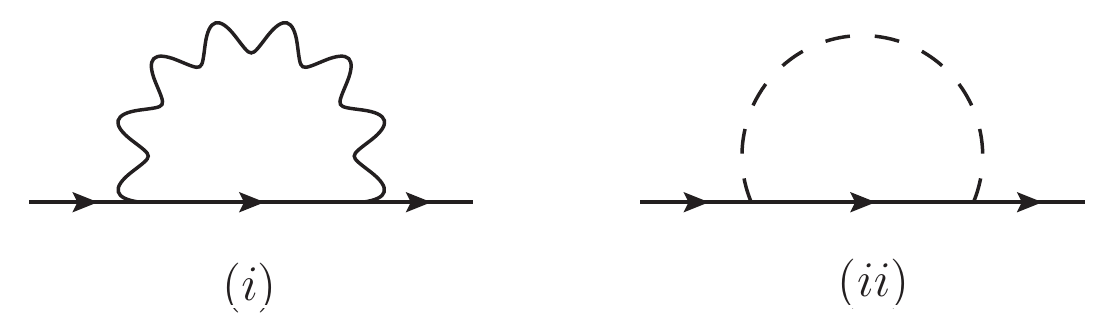}
\caption{Flavor-changing fermion self-energy diagrams corresponding to the blob in diagrams (a) and (b) of Figure~\ref{fig:vertices}.}
\label{fig:legs}
\end{figure}

The flavor-changing $G^\prime$ and $Z^\prime$ one-loop vertices are given in Figure~\ref{fig:vertices}, with the internal curvy (dashed) line denoting the $U_1$ LQ (Goldstone). Using the same normalization as in~\eqref{eq:effectiveVert}, the contribution from each diagram at $s=0$ and in the Feynman gauge reads
\begin{align}
V_{a+b}&=N_i\,\left[f_1(x_i)-x_i\,f_2(x_i)\right]\,,\no\\[2pt]
V_c&=g_\mathcal{V}^L\left[\frac{2\,x_i \ln x_i}{x_i-1}-f_1(x_i)\,c_i^2\right]+2\,g_\mathcal{V}^R\,f_3(x_i)\,,\no\\[2pt]
V_d&=-3\,g_{\mathcal{V}UU}\,f_1(x_i)\,,\no\\[2pt]
V_e&=g_\mathcal{V}^L\,x_i\,f_3(x_i)-g_\mathcal{V}^R\,x_i\left[f_2(x_i)+f_3(x_i)\right]\,,\no\\[2pt]
V_{f+g}&=-2\,g_{\mathcal{V}U\phi}\,f_1(x_i)\,,\no\\[2pt]
V_h&=g_{\mathcal{V}\phi\phi}\,x_i\,f_2(x_i)\,,
\end{align}
where $i=L\,(Q)$ for the quark (lepton) vertices, $x_{Q,L}=m_{Q,L}^2/m_U^2$, $N_L=3\,N_Q=3$, and
\begin{align}
f_1(x)&=\frac{x^2 \ln x}{(x-1)^2}-\frac{x}{x-1} \,,\no\\
f_2(x)&=\frac{1}{2}\left[\Delta_U+\frac{1}{2}-f_1(x)\right]\,,\no\\
f_3(x)&=\frac{x \ln x}{(x-1)^2}-\frac{x}{x-1} \,,
\end{align}
with $\Delta_U=\frac{1}{\epsilon}-\gamma_E+\ln 4\pi+\ln \frac{\mu^2}{m_U^2}$. Note that we have applied the unitarity relations discussed in~\ref{subsec:FCNCgen} in diagrams $(a.i)$, $(b)$, $(c)$ and $(d)$. Due to this unitarity cancellations, the contributions from these diagrams are finite. Diagrams $(f)$ and $(g)$ require a fermion mass insertion and are also finite. On the other hand, diagrams $(a.ii)$, $(e)$ and $(h)$ are divergent.

The couplings are given by
\begin{align}
g_{Z^\prime_q}^L&=3\,g_{Z^\prime_\ell}^L=3\,, &
g_{G^\prime}^L&=g_{G^\prime}^R=0\,,\no\\
g_{Z^\prime UU}&=4\,,&
g_{G^\prime UU}&=1\,,\no\\
g_{Z^\prime \phi U}&=4\left(1-x_{Z^\prime}\right)\,,&
g_{G^\prime \phi U}&=1-x_{G^\prime}\,,\no\\
g_{Z^\prime \phi\phi}&=4\left(1-\frac{x_{Z^\prime}}{2}\right)\,,&
g_{G^\prime \phi\phi}&=1-\frac{x_{G^\prime}}{2}\,,
\end{align}
The right-handed fermion couplings are different in model I and II. For model I, these couplings are zero, while in model II $g_{Z^\prime_q,Z^\prime_\ell}^L=g_{Z^\prime_q,Z^\prime_\ell}^R$.

\subsubsection{Contributions from radial modes}\label{app:RadialVert}

We discuss here the contributions from the radial modes, which we compute only for model I. The diagrams to be computed are the same as in Figure~\ref{fig:vertices} replacing the Goldstone by a radial leptoquark, except for $(h)$ which has two contributions: one with two radials, and one with a radial and a Goldstone. We find
\begin{align}
V_{a+b}^R&=-g_{hi}^2\,N_i\, x_i\,f_4(\tilde x_i)\,,\no\\
V_e^R&=g_\mathcal{V}^L\,g_{hi}^2\, x_i\,f_3(\tilde x_i)\,,\no\\
V_{f+g}^R&=2\,g_{\mathcal{V}Uh}\,g_{hi}\,x_i\,f_5(x_i,x_R)\,,\no\\
V_h^R&=g_{\mathcal{V}hh}\,g_{hi}^2\, x_i\,f_4(\tilde x_i)+g_{\mathcal{V}h\phi}\,g_{hi}\,x_i\,f_6(x_i,x_R)\,,
\end{align}
where $i=L\,(Q)$ for the quark (lepton) vertices, $\tilde x_i=m_i^2/m_{h_U}^2$, $x_R=m_{h_U}^2/m_U^2$, and the loop functions are defined as
\begin{align}
f_4(x)&=f_2(x)-\frac{1}{2}\ln x_R\,,\no\\
f_5(x_1,x_2)&=\frac{x_1 \ln x_1}{(x_2 - x_1) (x_1 - 1)} + \frac{x_2 \ln x_2}{(x_1 - x_2) (x_2 - 1)}\,,\no\\
f_6(x_1,x_2)&=\frac{1}{2}\left[\Delta_U+\frac{3}{2}+\frac{x_1^2 \ln x_1}{ (x_2 - x_1)(x_1 - 1)}\right.\no\\
&\quad\left.+\frac{x_2^2\ln x_2}{ (x_1 - x_2)(x_2 - 1)}\right]\,,
\end{align}
and the radial couplings are given by
\begin{align}
g_{hQ}&=\cot\beta\,,&
g_{hL}&=-\tan\beta\,,\no\\
g_{Z^\prime Uh}&=-4\sin\beta\cos\beta\,,&
g_{G^\prime Uh}&=2\sin\beta\cos\beta\,,\no\\
g_{Z^\prime \phi h}&=g_{Z^\prime Uh}\,,&
g_{G^\prime \phi h}&=g_{G^\prime Uh}\,,\no\\
g_{Z^\prime hh}&=1+2\sin^2\beta\,,&
g_{G^\prime hh}&=\cos^2\beta\,.
\end{align}
In the limit of very heavy radial mass compared to gauge boson and vector-like fermion masses, where $\tilde x_i\to0$ and $x_R\to\infty$, the loop functions above reduce to
\begin{align}
    f_3(\tilde x_i)&\to0\,,\no\\
    f_4(\tilde x_i)&\to\frac{1}{2}\left[\Delta_U-\ln x_R+\frac{1}{2}\right]\,,\no\\
    f_5(x_i,x_R)&\to0\,,\no\\
    f_6(x_i,x_R)&\to\frac{1}{2}\left[\Delta_U-\ln x_R+\frac{3}{2}\right]\,.
\end{align}

\subsubsection{Final result}\label{app:FCNCvertFinal}
Here we compile the results from the previous sections, using the same notation as in Section~\ref{subsec:FCNCVert}. For the gauge and Goldstone contributions, the regular functions in model I are
\begin{align}
F_{Z^\prime_q}^I(x_{Z^\prime},x_L,c_L)&=-x_{Z^\prime}F_1(x_L)+\frac{73 + 12\,c_L^2 - 3 x_L}{4 (x_L-1)}\no\\
&\quad-\frac{12 + (20 + 6\,c_L^2)\,x_L + 3 x_L^2}{2 (x_L-1)^2}\ln x_L\no\\
F_{Z^\prime_\ell}^I(x_{Z^\prime},x_Q,c_Q)&=-x_{Z^\prime}F_1(x_Q)+\frac{67 + 4 c_Q^2 - x_Q}{4 (x_Q - 1)}\no\\
&\quad-\frac{4 + 2 (14 + c_Q^2) x_Q + x_Q^2}{2 (x_Q-1)^2}\ln x_Q\no\\
F_{G^\prime}^I(x_{G^\prime},x_L,c_L)&=-\frac{x_{G^\prime}}{4}F_1(x_L)+\frac{4}{x_L-1}\no\\
&\quad-\frac{4}{(x_L - 1)^2}\,x_L\ln x_L\,,
\end{align}
while the regular functions in model II read
\begin{align}
F_{Z^\prime_q}^{II}(x_{Z^\prime},x_L,c_L)&=-x_{Z^\prime}F_1(x_L)+\frac{13 + 3\,c_L^2}{x_L - 1}\no\\
&\quad-\frac{13+3\,c_L^2}{(x_L-1)^2}\,x_L\ln x_L\,,\no\\
F_{Z^\prime_\ell}^{II}(x_{Z^\prime},x_Q,c_Q)&=-x_{Z^\prime}F_1(x_Q)+ \frac{15 + c_Q^2}{x_Q - 1}\no\\
&\quad-\frac{15 + c_Q^2}{(x_Q - 1)^2}\,x_Q\ln x_Q \,,\no\\
F_{G^\prime}^{II}(x_{G^\prime},x_L,c_L)&=F_{G^\prime}^I(x_{G^\prime},x_L,c_L)\,,
\end{align}
with $x_{Q,L}=m_{Q,L}^2/m_U^2$ and $c_{Q,L}$ as in~\eqref{eq:Omatrices}, and the function $F_1$ defined as
\begin{align}
F_1(x)=\frac{3 (x+5)}{2(x - 1)}-\frac{x+8}{(x - 1)^2}\,x\ln x\,.
\end{align}

The regular functions for the radial contributions, which we computed only for model I, are given by
\begin{align}
F_{Z^\prime_q}^{R_I}(x_{Z^\prime},x_R,\tilde x_L,x_L)&= F_2(x_R,\tilde x_L,x_L)\no\\
&\quad+\frac{(2 x_{Z^\prime} - 7)\,\tilde x_L}{(\tilde x_L - 1) (2 x_{Z^\prime} - 1)}\no\\
&\quad+\frac{6\tilde x_L
\ln \tilde x_L}{(\tilde x_L - 1)^2 (2 x_{Z^\prime} - 1)}\,,\no\\[2pt]
F_{Z^\prime_\ell}^{R_I}(x_{Z^\prime},x_R,\tilde x_Q,x_Q)&=F_2(x_R,\tilde x_Q,x_Q)\no\\
&\quad+\frac{(2 x_{Z^\prime}-5)\,\tilde x_Q}{(\tilde x_Q - 1) (2 x_{Z^\prime} - 3)}\no\\
&\quad+\frac{2\tilde x_Q
\ln \tilde x_Q}{(\tilde x_Q-1)^2 (2x_{Z^\prime}-3)}\,,\no\\[2pt]
F_{G^\prime}^{R_I}(x_R,\tilde x_L,x_L)&=F_2(x_R,\tilde x_L,x_L)+\frac{\tilde x_L}{\tilde x_L - 1}\,,
\end{align}
where we used the mass relations $x_{Z^\prime}=\frac{1}{2}+\sin^2\beta$ and $x_{G^\prime}=2\cos^2\beta$, and the function $F_2$ defined as
\begin{align}
F_2(x_1,x_2,x_3)&=-\frac{5}{2}+\frac{x_1^2+(9+x_3)\,x_1-x_3}{(x_1-x_3)(x_1-1)}\ln x_1\no\\
&\quad-\frac{x_2^2\,\ln x_2}{(x_2 - 1)^2}+\frac{2(x_3+4)\,x_3\ln x_3}{(x_3-x_1)(x_3-1)}\,.
\end{align}
Note that the singular points $x_{Z^\prime}\to3/2,1/2$ imply $\omega_{1,3}\to0$, respectively, for which $W\to\mathbb{1}$ and therefore vanishing FCNCs. 

\subsection{Box diagrams}\label{app:FCNCBoxes}

\begin{figure}[t]
\centering
\includegraphics[width=0.4\textwidth]{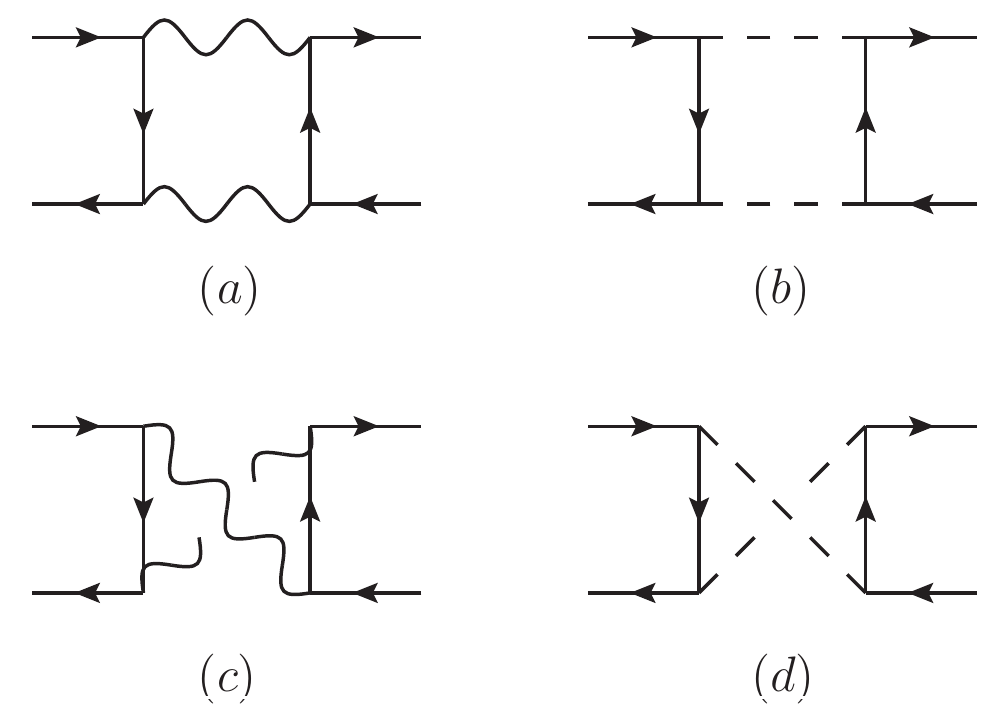}
\caption{Box diagrams contributing to the FCNCs.}
\label{fig:boxes}
\end{figure}

Here, we provide further details on the calculation of the box amplitudes. We present the results in the $SU(4)$ basis (see~\eqref{eq:SU4basis}), but we focus on the cases where only SM particles are present in the external states. There are four possible topologies contributing to these amplitudes. These are shown in Fig.~\ref{fig:boxes}, with the curvy line denoting the $U_1$ exchange, and the dashed line denoting a $\phi_U$ Goldstone exchange. Note that mixed diagrams with Goldstone and gauge leptoquarks necessarily contain a vector-like fermion as external state, which we do not consider here. Furthermore, we do not consider radial box contributions. These can be easily obtained from the Goldstone-box contribution by appropriately replacing the couplings, and are power-suppressed in the limit of heavy radial masses.

\subsubsection{Semileptonic amplitudes}

The amplitudes
for the semileptonic box contribution with two left-handed currents 
read\footnote{We define the amplitudes between an initial (partonic) state $|i\rangle$ and 
a final state $|f\rangle$  as 
$$
\mathcal{A}_{ij} = \langle f | T\{ e^{i
\int d^4 x {\mathcal L}_{\rm int} }
\} | i \rangle.
$$
}
\begin{align}
[\mathcal{A}^{q\ell}_{\rm NC}]^{ij\alpha\beta}=
i\, \frac{4G_U}{\sqrt{2}}\,\frac{\alpha_4}{4\pi}\,B_{q\ell }^{ij\alpha\beta}\,(\bar v_j \gamma_\mu P_L u_i)(\bar u_\alpha \gamma^\mu P_L v_\beta)\,.
\end{align}
Only the non-planar diagrams contribute to this amplitude, i.e. $B_{q\ell}=B_{q\ell}^c+B_{q\ell}^d$. The contributions of each diagram read
\begin{align}
[B_{q\ell}^c]^{ij\alpha\beta}&=2\,\delta_{ij}\delta_{\alpha\beta}\, B(0,0)\no\\
&+2\,\delta_{\alpha\beta}\,W_{i2}^*\,W_{j2}\,c_L^2\,[B(x_L,0)-B(0,0)]\no\\
&+2\,\delta_{ij}\,W_{2\alpha}^*\,W_{2\beta}\,c_Q^2\,[B(0,x_Q)-B(0,0)]\no\\
&+2\,W_{i2}^*\,W_{j2}\,W_{2\alpha}^*\,W_{2\beta}\,c_Q^2\,c_L^2\no\\
&\times[B(x_Q,x_L)-B(x_Q,0)-B(0,x_L)+B(0,0)]\,,\no\\[5pt]
[B_{q\ell}^d]^{ij\alpha\beta}&=\frac{1}{2}\,W_{i2}^*\,W_{j2}\,W_{2\alpha}^*\,W_{2\beta}\,x_Q\,x_L\,c_Q^2\,c_L^2\,\tilde B(x_Q,x_L)\,.
\end{align}
In the Feynman gauge, the loop functions are given by
\begin{align}\label{eq:BoxLoop}
B(x_1,x_2)&=\frac{1}{(1 - x_1) (1 - x_2)} + \frac{x_1^2 \ln x_1}{(1-x_1)^2 (x_1 - x_2)} \no\\
&\quad+ \frac{x_2^2 \ln x_2}{(1 - x_2)^2 (x_2 - x_1)}\,,\no\\[2pt]
\tilde B(x_1,x_2)&=\frac{B(x_1,x_2)}{4}\,.
\end{align}
which, as expected, are finite in this gauge. 

On the other hand, the box amplitudes for the case with one left-handed and one right-handed current are 
\begin{align}
[\mathcal{A}^{u(d)\ell}_{\rm NC}]^{ij\alpha\beta}=\frac{4G_U}{\sqrt{2}}\,\frac{\alpha_4}{4\pi}\,B_{u(d)\ell}^{ij\alpha\beta}\,(\bar v_j \gamma_\mu P_L u_i)(\bar u_\alpha \gamma^\mu P_R v_\beta)\,,\no\\
[\mathcal{A}^{qe}_{\rm NC}]^{ij\alpha\beta}=\frac{4G_U}{\sqrt{2}}\,\frac{\alpha_4}{4\pi}\,B_{qe}^{ij\alpha\beta}\,(\bar v_j \gamma_\mu P_L u_i)(\bar u_\alpha \gamma^\mu P_R v_\beta)\,.
\end{align}
In this case, there are no Goldstone contribution so the only relevant contribution comes from diagram (c) in Figure~\ref{fig:boxes}
\begin{align}
B_{u(d)\ell}^{ij\alpha\beta}&=\frac{1}{2}\,\delta_{ij}\delta_{\alpha\beta}\, B(0,0)\no\\
&+\frac{1}{2}\,\delta_{ij}\,W_{2\alpha}\,W_{2\beta}^*\,c_Q^2\,[B(0,x_Q)-B(0,0)]\,,\no\\
B_{qe}^{ij\alpha\beta}&=\frac{1}{2}\,\delta_{ij}\delta_{\alpha\beta}\, B(0,0)\no\\
&+\frac{1}{2}\,\delta_{\alpha\beta}\,W_{i2}^*\,W_{j2}\,c_L^2\,[B(x_L,0)-B(0,0)]\,.
\end{align}

\subsubsection{Hadronic amplitude}

The amplitude to the hadronic box contribution with two left-handed currents reads
\begin{align}
[\mathcal{A}^{qq}_{\rm NC}]^{ijkl}=  - i\,
\frac{4G_U}{\sqrt{2}}\,\frac{\alpha_4}{4\pi}\,[B_{qq}]^{ijkl}\,(\bar v_j \gamma_\mu P_L u_i)(\bar u_k \gamma^\mu P_L v_l)\,,
\end{align}
in which only planar diagrams contribute, i.e. $B_{qq}=B_{qq}^a+B_{qq}^b$. The contribution of each diagram is given by
\begin{align}
[B_{qq}^a]^{ijkl}&=\frac{1}{2}\delta_{ij}\delta_{kl}\, B(0,0)\no\\
&\quad+\frac{1}{2}\left(\delta_{kl}\,W_{i2}^*\,W_{j2}+\delta_{ij}\,W_{k2}\,W_{l2}^*\right)\,c_L^2\,\no\\
&\quad\times[B(x_L,0)-B(0,0)]\no\\
&\quad+\frac{1}{2}\,W_{i2}^*\,W_{j2}\,W_{k2}\,W_{l2}^*\,c_L^4\no\\
&\quad\times[B(x_L,x_L)-2\,B(x_L,0)+B(0,0)]\,,\no\\[5pt]
[B_{qq}^b]^{ijkl}&=\frac{1}{2}\,W_{i2}^*\,W_{j2}\,W_{k2}\,W_{l2}^*\,x_L^2\,c_L^4\,\tilde B(x_L,x_L)\,,
\end{align}
with the same loop functions as in the semileptonic case, c.f.~\eqref{eq:BoxLoop}. 

\subsubsection{Leptonic amplitude}

The corresponding box amplitude with two left-handed currents is given by
\begin{align}
[\mathcal{A}^{\ell\ell}_{\rm NC}]^{\alpha\beta\gamma\delta}= 
 - i
\frac{4G_U}{\sqrt{2}}\,\frac{\alpha_4}{4\pi}\,B_\ell^{\alpha\beta\gamma\delta}\,(\bar v_\beta \gamma_\mu P_L u_\alpha)(\bar u_\gamma \gamma^\mu P_L v_\delta)\,,
\end{align}
As in the hadronic case, only planar diagrams contribute. 
Each of them yields the following contribution
\begin{align}
[B_{\ell\ell}^a]^{\alpha\beta\gamma\delta}&=\frac{3}{2}\,\delta_{\alpha\beta}\delta_{\gamma\delta}\, B(0,0)\no\\
&\quad+\frac{3}{2}\left(\delta_{\gamma\delta}\,W_{2\alpha}\,W_{2\beta}^*+\delta_{\alpha\beta}\,W_{2\gamma}^*\,W_{2\delta}\right)\,c_Q^2\no\\
&\quad\times[B(x_Q,0)-B(0,0)]\no\\
&\quad+\frac{3}{2}\,W_{2\alpha}\,W_{2\beta}^*\,W_{2\gamma}^*\,W_{2\delta}\,c_Q^4\no\\
&\quad\times[B(x_Q,x_Q)-2\,B(x_Q,0)+B(0,0)]\,,\no\\[5pt]
[B_{\ell\ell}^b]^{\alpha\beta\gamma\delta}&=\frac{3}{2}\,W_{2\alpha}\,W_{2\beta}^*\,W_{2\gamma}^*\,W_{2\delta}\,x_Q^2\,c_Q^4\,\tilde B(x_Q,x_Q)\,,
\end{align}
with the loop functions in~\eqref{eq:BoxLoop}.

\subsection{SM dipole diagrams}\label{app:Dipoles}
\begin{figure}[t]
\centering
\includegraphics[width=0.4\textwidth]{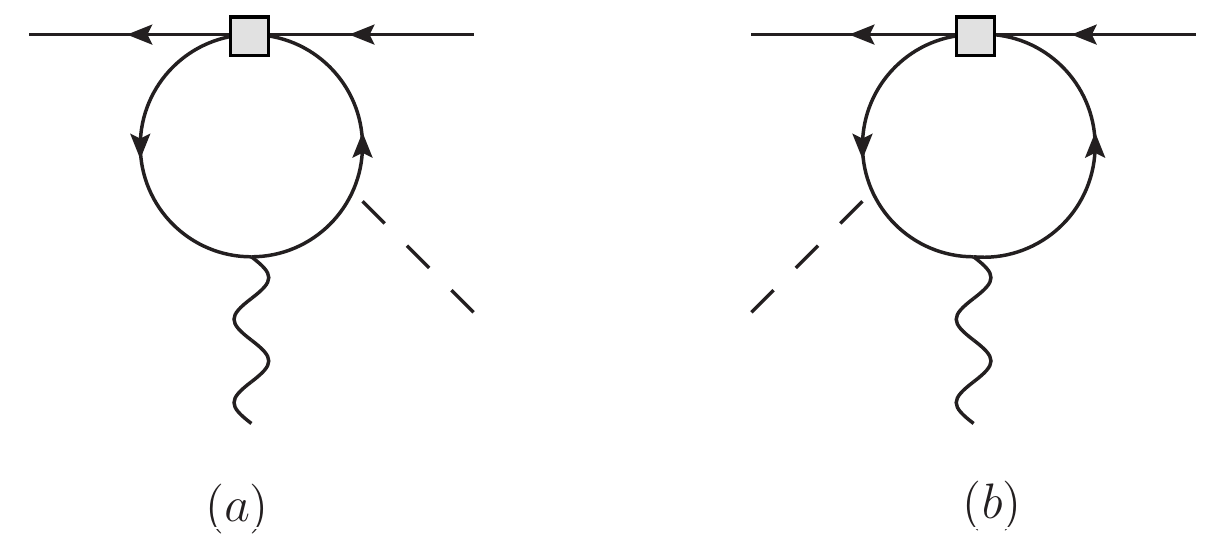}
\caption{Dipole diagrams in the SMEFT. The square denotes the insertion of the dimension-six operator resulting from integrating out the $U_1$ leptoquark. The dashed line represents the Higgs.}
\label{fig:EFTdipole}
\end{figure}

We provide here further details on the computation of the SM dipoles. The diagrams to be computed are (c), (d), (e) and (h) of Figure~\ref{fig:vertices} with the external gauge boson being a SM gauge boson, and with appropriate Higgs insertions in the fermion lines. The only cases where all diagrams are not vanishing are
the $B_{\mu\nu}$--dipoles, since both internal fermions and the LQ have non-vanishing $U(1)_Y$ charges. 
Considering the case of the $\cO^d_B$ operator as representative example and normalizing the Wilson coefficient as in \eqref{eq:Ldipoles}, we get
\bea
    C_{dB}&=& -\frac{y_b}{2} \left\{Y_\ell\, \left[D_c^L(x_L)+x_L\,D_e(x_L) \right] \right. \no\\
    &&\qquad\ +Y_U \left[D_d^L(x_L)+x_L D_h(x_L) \right] \big\}\no\\
    && -\frac{y_\tau}{2} \frac{ W_{21} }{   W_{12}^*W_{22} c_L^2 } \,\left[(Y_\ell+Y_e)\,(D_c^R-D_{\rm EFT})\right.\no\\
    &&\hspace{2.8cm}\left.+Y_U\,D_d^R \right]\,,
    \label{eq:CdB}
\eea
where $Y_{\ell,e,U}$ denote the corresponding hypercharges, and $D_k^{L,R}$ are the contributions from each of the diagrams in Figure~\ref{fig:vertices} evaluated in the Feynman gauge. In the limit of vanishing external momenta, these diagrams are given by
\begin{align}
D_c^L(x)&=x\,\frac{7 - 24 x + 21 x^2 - 4 x^3 + (6-12x) \ln x}{6 (x-1)^4}\,,\no\\[2pt]
D_d^L(x)&=-x \frac{5 - 15 x + 3 x^2 + 7 x^3 + 6(1 - 3 x) x\,\ln x}{
 12 (x-1)^4}\,,\no\\[2pt]
D_c^R&=-1\,,\qquad D_d^R=-\frac{3}{2}\,,\no\\[2pt] 
D_e(x)&=-\frac{2 + 3 x - 6 x^2 + x^3 + 6 x \ln x}{12 (x-1)^4}\,,\no\\[2pt]
D_h(x)&=-\frac{1 - 6 x + 3 x^2 + 2 x^3 - 6 x^2 \ln x}{12(x-1)^4}\,.
\end{align}
We further need to subtract the corresponding contributions from the EFT matrix elements. This is only non-vanishing for the diagrams associated to the $D_c^R$ contribution, shown in Figure~\ref{fig:EFTdipole}. We find
\begin{align}
D_{\rm EFT}=-1\,,
\end{align}
which exactly cancels the contribution from the corresponding UV diagram. This curious cancellation can swiftly be reproduced from computing the hard region of the corresponding loop graph in the full theory~\cite{Fuentes-Martin:2016uol,Beneke:1997zp,Smirnov:2002pj,Jantzen:2011nz} and seeing that it vanishes exactly.

The expression (\ref{eq:CdB}) can easily be matched to the general decomposition in \eqref{eq:WCdipoles}
defining the combined loop functions
\bea
G_{1} (x) &=& D_c^L(x)+x\, D_e(x)~,  \no\\
G_{2} (x) &=& D_d^L(x)+x\, D_h(x)~. 
\eea

\section{Details on the charged current computations}

\subsection{Two point function}\label{app:Uself}

The contribution of the vector-like fermions to the LQ 2-point function can be written as
\begin{equation}
    i \Sigma^{\mu\nu}_U(s) = i \frac{\alpha_4}{4\pi}\Sigma^{\tiny (f)}_U(s) g^{\mu\nu}\,,
\end{equation}
with the loop function $\Sigma_U(s)$ defined as
\bea
    \Sigma^{\tiny (f)}_U(s)&=& (N_f + N_{\chi_L})\, f_0(s) + N_{\chi_R}\, f_2(s,m_L,m_Q)\no\\
    && + 2N_{\chi_R}\,\Re(W_{22})\,c_L\, c_Q\,  \tilde{f_2}(s,m_L,m_Q)\no\\
    && + N_{\chi_L}\,|W_{22}|^2\, c_L^2\, c_Q^2\left[f_2(s,m_L,m_Q)\right. \no\\
    && - f_1(s,m_L) - f_1(s,m_Q)+f_0(s)\big]\no\\
    && + N_{\chi_L}\,c_Q\left[f_1(s,m_Q) - f_0(s)\right]\no\\
    && + N_{\chi_L}\,c_L\left[f_1(s,m_L) - f_0(s)\right]\,,
    \label{eq:Sigma}
\eea
where $N_f=4$ or $3$ depending on whether we include the right-handed neutrino in the loop, while $N_{\chi_L}=2$ and $N_{\chi_R} = 0\ (2)$ in model variant I (II). The $s$-dependent part of the loop functions is given by
\begin{align}
   \no f_0(s)&= -\frac{\text{$\Delta $}_U}{3}s+\frac{1}{3} s \log \left(-\frac{s}{m_U^2}\right)-\frac{5}{9}s\,,\\\no
    f_1(s,M)&=  -\frac{s}{3} \text{$\Delta $}_U -\frac{5 s}{9}+\frac{M^4}{6 s}+\frac{s}{3} \log \left(\frac{M^2}{m_U^2}\right)\\\no
    &\quad +\left(\frac{M^6}{6 s^2}-\frac{M^2}{2}+\frac{ s}{3}\right) \log \left(1-\frac{s}{M^2}\right)\,,\\\no
    f_2(s,M_1,M_2)&= -\frac{s}{3}\text{$\Delta $}_U -\frac{2 s}{9}+\left[\frac{1}{6s}\left(M_1^2-M_2^2\right)^2\right.\no\\
    &\quad+\frac{1}{6}\left(M_1^2+M_2^2\right)-\frac{s}{3}\Big] F\left(s,M_1^2,M_2^2\right)\no\\
    &\quad+\frac{s}{3}\frac{\left(M_1^2 \log \left(\frac{M_1^2}{m_U^2}\right)-M_2^2 \log \left(\frac{M_2^2}{m_U^2}\right)\right)}{M_1^2-M_2^2}\,,\no\\
    \tilde{f_2}(s,M_1,M_2)&= - M_1 M_2 F\left(s,M_1^2,M_2^2\right)\,,
\end{align}
with $F\left(s,M_1^2,M_2^2\right)$ as reported in~\cite{Bohm:1986rj}. The loop functions defined above contain constant divergent pieces that can be absorbed into the definition of the physical mass. Employing the on-shell renormalization scheme as in~\cite{Fuentes-Martin:2019ign}, with degenerate vector-like fermion masses equal to the LQ mass $m_L=m_Q=m_U$, their effect to the 2-point function at $s=0$ has the form
\bea\label{eq:dSigma0}
    \delta\Sigma^{\tiny (f)}_U(0)&=& - \frac{1}{3}(N_f + N_{\chi_L}) + \frac{3}{8}\,N_{\chi_L}(c_L+c_Q)\no\\
    && - \frac{1}{2}\,N_{\chi_L}\left(1+\frac{2}{9}\sqrt{3}\pi\right)|W_{22}|^2\,c_L^2 c_Q^2\no\\
    && - \frac{2}{9}\,N_{\chi_R}\left(\sqrt{3}\pi - 6\right)\Re(W_{22})\,c_L c_Q \no\\
    && + \frac{1}{9}\,N_{\chi_R}\left(5\sqrt{3}\pi-27\right) \,.
\eea
Setting $c_Q=c_L=W_{22}=1$ and $N_{\chi_L}=2$, the finite correction is
\begin{equation}
    \delta\Sigma^{\tiny (f)}_U(0)+\frac{N_f}{3} = 
    \begin{cases} 
         -\frac{1}{6}-\frac{2\sqrt{3} \pi }{9} \approx -1.04\,,& N_{\chi_R} = 0\,, \\[2pt]
         \frac{19}{6}-\frac{8\sqrt{3} \pi }{9} \approx -1.67\,,& N_{\chi_R} = 2\,.
    \end{cases}
\end{equation}
Here, we have isolated the effect of the vector-like fermions, i.e. we have removed the $-N_f/3$ factor corresponding to the SM fields. The correction from adding the vector-like fermions tends to decrease the low-energy enhancement at NLO calculated in~\cite{Fuentes-Martin:2019ign}. In particular, for fixed on-shell coupling $g_4=3$, we find an $\mathcal{O}(5\%-10\%)$ reduction of the Wilson coefficients studied in~\cite{Fuentes-Martin:2019ign}.

\subsection{Charged current box contributions}\label{app:CCBoxes}

\begin{figure}[t]
\centering
\includegraphics[width=0.35\textwidth]{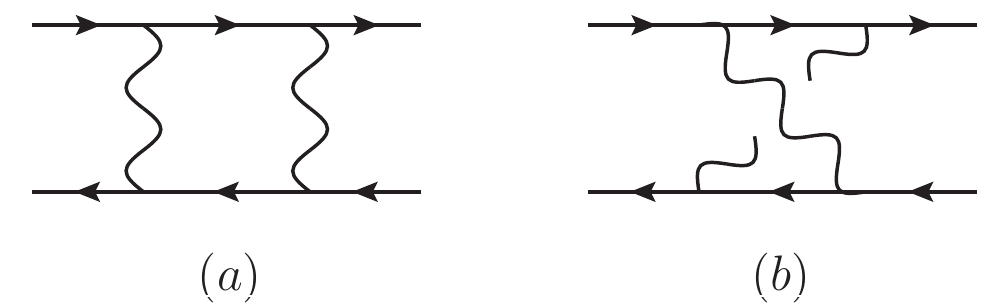}
\caption{Charged-current box diagrams.}
\label{fig:boxesCC}
\end{figure}

In analogy to the neutral-current boxes, also for the charged-current boxes we present the results in the $SU(4)$ basis. We focus on the $\mathcal{Q}_2 \to \mathcal{Q}_1$ flavor-violating amplitude that contributes to $b\to c\tau\nu$ transitions:
\begin{align}
[\mathcal{A}^{q\ell}_{\rm CC}]^{1211}&=-i\,\frac{4G_U}{\sqrt{2}}\,\frac{\alpha_4}{4\pi}\,W_{11}^*\,W_{21} B^{1211}_{L,R}\nonumber\\
&\quad\times(\bar v_{\mathcal{Q}_2} \gamma_\mu P_{L,R}\, u_{\mathcal{Q}_1})(\bar u_{\mathcal{L}_1} \gamma^\mu P_L v_{\mathcal{L}_1})\,.
\end{align}
In this case, diagrams with Goldstone boson exchange do not appear, and there are two possible topologies contributing to these amplitudes, namely diagrams (a) and (b) in Figure~\ref{fig:boxesCC}. We decompose the different contributions as
\begin{align}
B^{1211}_{L,R}&=\sum_{i,V_1,V_2}B_{L,R}^{i,V_1,V_2}\,, \no\\
& \equiv B^{1111}_{L,R}(0) + c^2_Q\, \Delta B_{L,R}^{1211} (x_Q)~,
\label{eq:C8}
\end{align}
where $i=a,b$ denotes the topology of the box, and $V_{1,2}=U,Z^\prime,G^\prime$ indicate the gauge bosons in the propagators. The separate contributions of the various diagrams are
\begin{align}
B^{a,UZ^\prime}_L=\frac{B^{a,UZ^\prime}_R}{4}&=\frac{3}{8}\,\bar B (x_{Z^\prime},0)\,,\nonumber\\
B_L^{a,Z^\prime U}=\frac{B_R^{a,Z^\prime U}}{4}&=\frac{1}{24}\left\{\bar B(x_{Z^\prime},0))\right.\nonumber\\
&\quad\left.+c_q^2\left[\bar B(x_{Z^\prime},x_Q-\bar B(x_{Z^\prime},0)\right]\right\}\,,\nonumber\\
B_L^{a,G^\prime U}=\frac{B_R^{a,G^\prime U}}{4}&=\frac{4}{3}\left\{\bar B(x_{G^\prime},0)\right.\nonumber\\
&\quad\left.+c_q^2\left[\bar B(x_{G^\prime},x_Q)-\bar B(x_{G^\prime},0)\right]\right\}\,,\nonumber\\
B_L^{b,UZ^\prime}=4B_R^{b,UZ^\prime}&=\frac{1}{2}\,\bar B(x_{Z^\prime},0)\,,\nonumber\\
B_L^{b,Z^\prime U}=4B_R^{b,Z^\prime U}&=\frac{1}{2}\left\{\bar B(x_{Z^\prime},0)\right.\nonumber\\
&\quad\left.+c_q^2\left[\bar B(x_{Z^\prime},x_Q)-\bar B(x_{Z^\prime},0)\right]\right\}\,,
\end{align}
with the loop function defined as
\begin{align}
\bar B(x_1,x_2)&=\frac{x_1 \ln x_1}{(x_1-1) (x_1 - x_2)} + \frac{
 x_2 \ln x_2}{(x_2-1) (x_2-x_1)}\,.
\end{align}
As indicated in (\ref{eq:C8}), summing all contributions we can decompose the result into a term independent from the vector-like mass, which is equivalent to the loop function appearing in the flavor-conserving amplitude, and a term which vanishes in the limit $x_Q \to 0$. The former coincides with the loop function obtained in~\cite{Fuentes-Martin:2019ign}. Using the notation of the latter paper, we have
\bea
B^{1111}_{L}(0) &=& \frac{4}{3} f_{G^\prime} + \frac{17}{12} f_{Z^\prime}~, \no\\ 
B^{1111}_{R}(0) &=& \frac{16}{3} f_{G^\prime} + \frac{23}{12} f_{Z^\prime}~,
\eea
with $f_{V}=\ln x_V/(x_V-1)$ and $x_V=m_V^2/m_U^2$. In the custodial limit for the massive vectors, the terms that vanish at $x_Q\to 0$ read
\begin{align}
\Delta B^{1211}_L (x_Q) =  \frac{15}{8}\,\frac{x_Q}{(1-x_Q)^2}\left(1-x_Q+\ln x_Q\right)\,,\no\\
\Delta B^{1211}_R (x_Q) =  \frac{45}{8}\,\frac{x_Q}{(1-x_Q)^2}\left(1-x_Q+\ln x_Q\right)\,.
\end{align}


\bibliographystyle{JHEP}
\bibliography{references}

\end{document}